\documentclass[10pt,aps,floatfix,amssymb,prd,twocolumn,superscriptaddress,nofootinbib,preprintnumbers]{revtex4-2}
	
\usepackage{ragged2e}
\usepackage{amssymb,amsmath,verbatim,mathtools,needspace,enumitem,etoolbox,graphicx,microtype,afterpage,bm}
\usepackage[dvipsnames]{xcolor}
\usepackage[T1]{fontenc}
\usepackage[utf8]{inputenc}

\usepackage[normalem]{ulem}

\definecolor{linkcolor}{rgb}{0.0,0.3,0.5}
\usepackage{booktabs}
\usepackage[unicode, colorlinks=true, linkcolor=linkcolor, citecolor=linkcolor, filecolor=linkcolor,urlcolor=linkcolor, pdfusetitle]{hyperref}
\usepackage{tabularx}

\usepackage{multirow}
\usepackage{relsize}

\newcommand{\nn}{\nonumber}

\renewcommand\({\left(}
\renewcommand\){\right)}
\renewcommand\[{\left[}
\renewcommand\]{\right]}

\newcommand{\ra}{\rightarrow}

\def\lsim{\raise 0.4ex\hbox{$<$}\kern -0.8em\lower 0.62
ex\hbox{$\sim$}}

\def\gsim{\raise 0.4ex\hbox{$>$}\kern -0.8em\lower 0.62
ex\hbox{$\sim$}}

\def\lbar{{\hbox{$\lambda$}\kern -0.7em\raise 0.6ex
\hbox{$-$}}}

\newcommand\eq[1]{Eq.~(\ref{#1})}
\newcommand\eqs[2]{Eqs.~(\ref{#1}) and (\ref{#2})}
\newcommand\Eq[1]{Equation~(\ref{#1})}

\newcommand\eqss[3]{Eqs.~(\ref{#1}), (\ref{#2}) and (\ref{#3})}

\newcommand\p{\partial}

\newcommand\ee{\end{equation}}
\newcommand\be{\begin{equation}}
\def\bea{\begin{array}}
\def\eea{\end{array}}\def\ea{\end{array}}
\newcommand\ees{\end{eqnarray}}
\newcommand\bees{\begin{eqnarray}}
\def\nn{\nonumber}

\def\f{\phi}

\def\dslash{\hspace{-1mm}\not{\hbox{\kern-2pt $\partial$}}}
\def\Dslash{\not{\hbox{\kern-2pt $D$}}}
\def\pslash{\not{\hbox{\kern-2.1pt $p$}}}
\def\kslash{\not{\hbox{\kern-2.3pt $k$}}}
\def\qslash{\not{\hbox{\kern-2.3pt $q$}}}

\newcommand{\vx}{{\bf x}}

\newcommand{\bdot}{{\bf\cdot}} 

\def\p1{{\bf p}_1}
\def\p2{{\bf p}_2}
\def\k1{{\bf k}_1}
\def\k2{{\bf k}_2}

\newcommand{\hatn}{\hat{\bf n}}

\newcommand{\dddM}{\kern 0.2em \raise 1.9ex\hbox{$...$}\kern -1.0em \hbox{$M$}}
\newcommand{\dddQ}{\kern 0.2em \raise 1.9ex\hbox{$...$}\kern -1.0em \hbox{$Q$}}
\newcommand{\dddI}{\kern 0.2em \raise 1.9ex\hbox{$...$}\kern -1.0em\hbox{$I$}}
\newcommand{\dddJ}{\kern 0.2em \raise 1.9ex\hbox{$...$}\kern-1.0em
\hbox{$J$}}
\newcommand{\dddcalJ}{\kern 0.2em \raise 1.9ex\hbox{$...$}\kern-1.0em
\hbox{${\cal J}$}}

\newcommand{\dddO}{\kern 0.2em \raise 1.9ex\hbox{$...$}\kern -1.0em
\hbox{${\cal O}$}}
\def\dddz{\raise 1.5ex\hbox{$...$}\kern -0.8em \hbox{$z$}}
\def\dddd{\raise 1.8ex\hbox{$...$}\kern -0.8em \hbox{$d$}}
\def\dddbd{\raise 1.8ex\hbox{$...$}\kern -0.8em \hbox{${\bf d}$}}
\def\ddbd{\raise 1.8ex\hbox{$..$}\kern -0.8em \hbox{${\bf d}$}}
\def\dddx{\raise 1.6ex\hbox{$...$}\kern -0.8em \hbox{$x$}}

\newcommand{\hti}{\tilde{h}}

\newcommand{\msun}{M_{\odot}}

\newcommand{\nab}{\langle\tilde{n}^{{\rm eff},*}_a(f)\tilde{n}^{\rm eff}_b(f)\rangle}
\newcommand{\neff}{\langle\tilde{n}^{\rm eff, *}_a\, \tilde{n}^{\rm eff}_b\rangle}
\newcommand{\nev}{N_{\rm ev}}

\newcommand{\inner}[2]{\left(\, #1\,\, ,\, #2\,\right)_{ab}}
\newcommand{\scalarp}[2]{\left(#1 \Big| #2\right)_a}
\newcommand{\FM}{\Gamma}
\newcommand{\CM}{\Sigma}
\newcommand{\de}[1]{\partial_{#1}}
\newcommand{\likelihood}[2]{\mathcal{L}\left(#1\big|#2\right)}

\newcommand{\ft}[1]{\tilde{#1}}

\newcommand{\ftconj}[1]{\ft{#1}^{*}}

\newcommand{\snr}{\rm{SNR}}
\newcommand{\snrth}{{\rm SNR}_{\rm th}}
\newcommand{\snrobs}{{\rm SNR}^{\rm obs}}
\newcommand{\snrpair}{\(\frac{S_{\cal N}}{N}\)_{ab}^{\rm max}}
\newcommand{\snconf}{(S_{\cal N}/N)_{ab}^{\rm  max}} 

\newcommand{\Seffab}{S_{ab}^{\rm  eff}}
\newcommand{\omerr}{\Omega^{\rm err}_{ab}}

\newcommand{\omgwa}{\Omega_{\rm gw}^{\rm astro}}

\newcommand{\omgw}{\Omega_{\rm gw}}

\newcommand{\oeff}{\Omega_{ab}^{\rm  eff}}

\newcommand{\Snab}{S_{n,ab}}
\newcommand{\hnab}{h_{n,ab}}
\newcommand{\hcab}{h_{c,ab}}
\newcommand{\nnoise}{N_n}
\newcommand{\parset}{\bm{\theta}}
\newcommand{\mlpar}{\bm{\theta}^{\rm ML}}
\newcommand{\obspar}{\bm{\theta}^{\rm obs}}
\newcommand{\trpar}{\bm{\theta}^{\rm true}}

\newcommand{\normdist}[1]{\mathcal{N}\left(0, #1\right)}

\allowdisplaybreaks
\usepackage{tikz}
\usetikzlibrary{shapes.misc, positioning, arrows.meta}
\usepackage{framed}
\usepackage{hyperref}
\usepackage[capitalise]{cleveref}
\usepackage{nicefrac}

\hypersetup{colorlinks, citecolor=linkcolor, linkcolor=linkcolor, urlcolor=blue}

\crefname{section}{Sec.}{Sec.}
\Crefname{section}{Section}{Section}
\crefname{appendix}{App.}{App.}
\Crefname{appendix}{Appendix}{Appendix}
\crefname{footnote}{footnote}{footnote}
\Crefname{footnote}{footnote}{footnote}
\crefname{enumi}{item}{item}
\Crefname{enumi}{Item}{Item}
\crefname{page}{page}{page}
\Crefname{page}{Page}{Page}

\usepackage{siunitx}
\DeclareSIUnit \parsec {pc}
\DeclareSIUnit \arcsecondfull {arcsec}
\DeclareSIUnit \year{yr}
\DeclareSIUnit \day{day}
\DeclareSIUnit \hour{hr}
\DeclareSIUnit \radiant{rad}
\DeclareSIUnit \degfull{deg}
\DeclareSIUnit \erg {erg}
\DeclareSIUnit \Lsun {L_\odot}
\DeclareSIUnit \Msun {M_\odot}
\DeclareSIUnit \AstroUnit {au}

\renewcommand{\Re}{{\rm Re}}
\renewcommand{\Im}{{\rm Im}}

\newcommand{\unige}{D\'epartement de Physique Th\'eorique, Universit\'e de Gen\`eve, 24 quai Ernest Ansermet, 1211 Gen\`eve 4, Switzerland}

\newcommand{\gwsc}{Gravitational Wave Science Center (GWSC), Universit\'e de Gen\`eve, CH-1211 Geneva, Switzerland}

\newcommand{\mrs}{Aix-Marseille Universit\'e, Universit\'e de Toulon, CNRS, CPT, Marseille, France}

\begin{document}

\title{Confusion noise from astrophysical  backgrounds  at third-generation \texorpdfstring{\\}{} gravitational-wave detector networks}

\author{Enis Belgacem}
\email{enis.belgacem@unige.ch}
\affiliation{\unige}
\affiliation{\gwsc}

\author{Francesco Iacovelli}
\email{francesco.iacovelli@unige.ch}
\affiliation{\unige}
\affiliation{\gwsc}

\author{Michele Maggiore}
\email{michele.maggiore@unige.ch}
\affiliation{\unige}
\affiliation{\gwsc}

\author{Michele Mancarella}
\email{mancarella@cpt.univ-mrs.fr}
\affiliation{\mrs}

\author{Niccol\`o Muttoni}
\email{niccolo.muttoni@unige.ch}
\affiliation{\unige}
\affiliation{\gwsc}

\date{\today}

\begin{abstract}
At   third-generation (3G)  gravitational-wave detector networks,  compact binaries coalescences   produce a ``confusion noise'' due to unresolved sources and to the error in the reconstruction  of  resolved sources, that can  degrade the sensitivity to cosmological backgrounds. We show how to characterize
from first-principles this astrophysical confusion noise by reconstructing  the resolved sources at a detector network, subtracting them from the data stream of each detector of the network,  and then computing 
the correlation among  detector pairs of these ``partially cleaned''  data streams that, beside instrumental noise, contain the unresolved sources and the error on the resolved sources.  In a two-detector correlation, this residual astrophysical background then acts as an effective correlated noise. We point out that its effect, in the search for a cosmological background, must  be evaluated  by correlating two detectors with the filter function that optimizes the given cosmological search (and not the search for the astrophysical background itself) and therefore depends on the specific cosmological signal that we want to extract from the data. 
We show how to obtain search-independent upper bounds on the effect of this astrophysical confusion noise,  and  how to characterize its actual effect on a given cosmological search.
We then apply this methodology to an example of a  3G network and, for a realistic population of mergers,  we evaluate explicitly  the upper bound on the effect of the astrophysical confusion noise, as well as its actual effect in the search of selected examples of power-law cosmological backgrounds and of backgrounds featuring broad peaks, as in  phase transitions.   
\end{abstract}


\maketitle

{
  \hypersetup{linkcolor=black}
  \tableofcontents
}
\hypersetup{linkcolor=linkcolor}

\section{Introduction}\label{intro}

After the first direct observation of gravitational waves (GWs) emitted by a compact binary coalescence~\cite{LIGOScientific:2016aoc}, the LIGO-Virgo Collaboration (LVC), later joined by KAGRA (LVK), has performed  three observing runs, which have led to the detection of about 90  binary black hole (BBH) coalescences,  two binary neutron star (BNS) coalescences, and two neutron star--black hole (NSBH) coalescences~\cite{LIGOScientific:2020ibl,KAGRA:2021vkt}. A fourth observational run is currently ongoing. Despite the fact that, at the level of sensitivity of the third observing run, BBHs were detected at a rate of a few per week, the events currently observed, as well as those that will be detected with upgrades of the current (second-generation) detectors, only constitute the tip of the iceberg: the most recent population synthesis models, combined with the LVK observations,
predict that there are ${\cal O}(10^5)$ BBHs and BNSs coalescing per year in the Universe~\cite{KAGRA:2021duu,Mapelli:2017hqk, Giacobbo:2017qhh, Giacobbo:2018hze, Fishbach:2018edt, Rodriguez:2018rmd, Vitale:2018yhm, Mapelli:2019ipt, Baibhav:2019gxm, Bavera:2020uch, Santoliquido:2020axb, Mapelli:2021gyv, Broekgaarden:2021efa, vanSon:2021zpk, Chu:2021evh, COMPASTeam:2021tbl, Chruslinska:2022ovf}, corresponding to one event every few minutes.

In contrast, third-generation (3G) detectors such as Einstein Telescope (ET)~\cite{Hild:2008ng,Punturo:2010zz,Hild:2010id,Maggiore:2019uih} and  Cosmic Explorer (CE)~\cite{Reitze:2019iox,Evans:2021gyd,Evans:2023euw} will detect the vast majority of these events~\cite{Sathyaprakash:2012jk,Maggiore:2019uih}. For instance, ET alone
would detect basically $100\%$ of the population up to $z\sim 1$, and about $67\%$ of the BBH population out to $z\sim 20$, while a network of ET with two CEs would detect about $100\%$ of the BBH population out to $z\simeq2$, and $93\%$  up to $z\sim 20$; see  Refs.~\cite{Iacovelli:2022bbs,Borhanian:2022czq} for   comprehensive studies  of the ET capabilities  for coalescing binaries, and  
Ref.~\cite{Branchesi:2023mws} for a  discussion of the science case of ET, including a comparison between a single triangular design and a design with two widely separated L-shaped detectors (see also \cite{Puecher:2023twf,Bhagwat:2023jwv,Franciolini:2023opt,Iacovelli:2023nbv} for further follow-up studies,  and Refs.~\cite{Gupta:2023lga,Iacovelli:2024mjy} for  studies of  other 3G network configurations).

While extremely interesting in itself, the superposition of all these signals of astrophysical origin is a foreground that could  mask a stochastic GW background of cosmological origin. In general,  there are two sources of confusion noise. One is the stochastic background due to the unresolved astrophysical sources, i.e. those that are below the detection threshold. The second is due to sources above the detection threshold: their contribution to the detector strain can in principle be subtracted but, unavoidably, there will be an error in the  reconstruction of the source  parameters. The errors in the reconstruction of the resolved sources add up, and generate a second source of confusion noise, that might obstruct the access to  cosmological backgrounds~\cite{Cutler:2005qq,Harms:2008xv,Regimbau:2016ike,Pan:2019uyn,Sachdev:2020bkk,Sharma:2020btq,Perigois:2020ymr,Lewicki:2021kmu,Perigois:2021ovr,Zhou:2022otw,Zhou:2022nmt,Zhong:2022ylh,Pan:2023naq,Zhong:2024dss,Li:2024iua}.

The  aim of this paper is to compute from first principles  the effect of these astrophysical backgrounds on the search for cosmological backgrounds. 
Stochastic GW backgrounds, at ground-based detectors, can only be searched by correlating the outputs of different detectors. Our strategy will  be to generate an ensemble of  BBH and BNS signals from a realistic population model, reconstruct the source parameters of the resolved sources at a 3G detector network in a given noise realization (the specific noise realization will also determine which sources are resolved and which are not), and subtract the reconstruction of the resolved  sources
from  the output stream of each detector in the network. The output of each detector will at this point contain  a specific realization of instrumental noise, the signals  from the sources that we have injected  that, in the given noise realization, are unresolved, and, for the resolved sources,  the difference between the actual injected signals and the reconstructed ones (plus, possibly, a cosmological signal).
Both  the  unresolved sources and the  errors in the subtraction of resolved sources will  generate an effective correlated noise in the output of a detector pair.
We will show how to characterize this correlated  component of the noise and we will compute it, averaging the results over a large number of noise realizations.  We will then be able to evaluate how it affects the sensitivity to cosmological stochastic backgrounds of 3G detector networks. 

The paper is organized as follows.
In \cref{sect:astroBKG} we recall some standard definitions for characterizing stochastic backgrounds, in order to fix the notation, paying attention in particular  to   the case of backgrounds generated by the superposition of a discrete number of astrophysical sources. 

In \cref{sect:cross} we  examine matched filtering for stochastic backgrounds. After recalling standard results for continuous backgrounds in the simplest version (uncorrelated noise and stationary, unpolarized, isotropic and Gaussian GW background), we will  discuss some extensions relevant to the subtraction of the astrophysical background. In particular, we will discuss  the extension to correlated noise (since we will eventually treat the residual astrophysical background as a correlated noise).

In \cref{sect:Astrocorrelated noise} we  show how  the 
astrophysical background due to unresolved sources, and to the error in the reconstruction of resolved sources, generates a correlated noise in the output of detector pairs; we will  show how to characterize and evaluate it,  
posing the conceptual basis for a first-principle evaluation of the effect of this astrophysical ``confusion noise''. 

In \cref{sect:sourcepar_reconstruction} we discuss our procedure for the reconstruction of the parameters of the injected signals, which (even if we deal with a large population of signals) goes beyond the Fisher-matrix approach, and takes into account the effect of the specific noise realization.

In \cref{sect:rescorr} we will  present our numerical results for the subtraction of the BBH and BNS backgrounds, at a 3G detector network;
as an example, we will consider a network made by ET in its \SI{10}{\kilo\meter} triangular configuration plus 2 CE detectors, of \SI{40}{\kilo\meter} and \SI{20}{\kilo\meter}, respectively, which is a standard benchmark used in the literature in this context.

Finally, \cref{sect:conclusions} contains our Conclusions, while in 
\cref{sect:comparison} we compare our results with some recent works in the literature.

\section{Characterization of   stochastic  backgrounds}\label{sect:astroBKG}

Stochastic GW backgrounds are given by the incoherent superposition of signals that reach the detectors on Earth. These can have an astrophysical origin (e.g. unresolved compact binary coalescences, or supernovae) or a cosmological origin (i.e. phenomena happening in the early Universe such as cosmic strings, phase transitions, reheating, etc);  see Refs.~\cite{Maggiore:1999vm,Maggiore:2007ulw,Regimbau:2011rp,Romano:2016dpx,Maggiore:2018sht,Caprini:2018mtu,Christensen:2018iqi,Lawrence:2023buo} for reviews.
In this section we  recall some  standard definitions, also in order to fix the notation.

The output of a given GW detector, labeled by an index $a$, in presence of a GW signal, has the form 
\be\label{sequalnplush}
s_a(t) =n_a(t)+h_a(t)\, ,
\ee
where $n_a(t)$ is the noise in the $a$-th detector and $h_a(t)$ is  a generic stochastic GW signal (projected onto the $a$-th detector),  that we aim to  extract  from the data.

The properties of the noise and those of a stochastic signal can only be characterized statistically, i.e. in terms of ensemble averages. One  must, however, distinguish between two different ensemble averages. In the presence of detector noise, we will need to perform an ensemble average of various quantities over the possible realizations of the noise. 
In contrast, for GW signals,  
the ensemble average is, conceptually, an average over many different realizations of the Universe. Of course, in practice we only have a single realization of the Universe. However, when the signal is a superposition of a virtually infinite number of contributions (as, e.g., the stochastic GW background due to the amplification of vacuum fluctuations in the early Universe), the signal received is perfectly well described by such a fictitious average over different realizations of the Universe. In contrast, for the superposition of astrophysical signals, there can be  some  dependence on the specific realization. In the following, we will use the notation 
$\langle \ldots \rangle_n$
for the average over noise realizations, and 
$\langle \ldots \rangle_U$
for the average over different realizations of the Universe. 
So,  for a quantity that depends both on the noise and on the GW stochastic signal, 
\be\label{defaveragenU}
\langle \ldots \rangle \equiv \langle \hspace{3mm} \langle \ldots \rangle_n \hspace{3mm} \rangle_U \, .
\ee
It is convenient to work in Fourier space, in terms of the Fourier modes $\tilde{n}_a(f)$ and $\tilde{h}_a(f)$.
If the noise is stationary, the correlator $\left\langle \tilde{n}_a^*(f)\tilde{n}_a(f') \right\rangle_n $  must be proportional to $\delta(f-f')$, so we can write
\be\label{Sn1noT}
\left\langle \tilde{n}_a^*(f)\tilde{n}_a(f') \right\rangle_n =
\delta(f-f')\frac{1}{2}S_{n,a}(f)\, .
\ee
The function $S_{n,a}(f)$ defined by this equation is called the noise power spectral density, PSD.
Actually, \eq{Sn1noT} is an idealization that assumes that the noise average is taken 
over an infinite observation time. In a finite observation time $T$,  rather than a Dirac delta $\delta(f-f')$ we must use a regularized Dirac delta, i.e.
\be\label{Sn1}
\left\langle \tilde{n}_a^*(f)\tilde{n}_a(f') \right\rangle_n =
\delta_T(f-f')\frac{1}{2}S_{n,a}(f)\, ,
\ee
where
\be\label{defdeltaT}
\delta_T(f-f')=\int_{-T/2}^{T/2} dt\, e^{-2\pi i (f-f') t}\, .
\ee
Observe that, setting $f=f'$,   
\be\label{deltazero}
\delta_T(0) =T\, .
\ee
Without loss of generality, we can always assume that 
$\langle n_a(t)\rangle_n =0$. If, furthermore, the noise is Gaussian, all information about it is contained in the two-point correlator $\langle \ft{n}_a^*(f)\ft{n}_a(f') \rangle_n $, so a Gaussian and stationary noise is characterized uniquely by its PSD. 

Taking the complex conjugate of \eq{Sn1} we get that $S_{n,a}(f)$ is a real function, while from the reality of $n_a(t)$ it follows that $\ft{n}_a^*(f)=\ft{n}_a(-f)$, so the PSD satisfies
\be\label{Snreality}
S^*_{n,a}(f)=S_{n,a}(f)\, ,\qquad
S_{n,a}(-f)=S_{n,a}(f)\, .
\ee
Let us characterize next the statistical properties of a stochastic GW signal. A continuous superposition of GWs coming from all directions, with propagation directions labeled by a unit vector $\hatn$,  can be written as 
\bees\label{snrhab}
&&\hspace*{-14mm}h_{kl}(t,\vx)
=\int_{-\infty}^{\infty} df \int_{S^2} d^2\hatn\,\nn\\
&&\hspace*{-4mm}\times\hspace*{0mm}\sum_{A=+,\times}  \hti_A(f,\hatn) e^A_{kl}(\hatn) \, \,   e^{-2\pi i f(t-\hatn\cdot\vx /c )}
\, ,
\ees
where $k,l=1,2,3$ are spatial indices (which, in the TT gauge, reduce to  indices taking two values,  in the transverse plane), $A=+,\times$ labels the two polarizations, and $e_{kl}^{A}$ are the polarization tensors, normalized as 
\be\label{normaee}
e_{kl}^{A}(\hatn)e_{kl}^{A'}(\hatn)=2\delta^{AA'}\, ,
\ee
where the sum over the spatial indices $k,l$ is understood
(we follow the notation and conventions in Ref.~\cite{Maggiore:2007ulw}). 
In a given detector,  located at the position $\vx_a$, the corresponding observed signal $h_a(t)$  is obtained convolving each polarization with functions that encode the detector response,
\bees\label{hFhF}
&&\hspace*{-14mm}h_a(t)= \int_{-\infty}^{\infty} df \int_{S^2} d^2\hatn\,\nn\\
&&\hspace*{-4mm}\times\hspace*{0mm} \sum_{A=+,\times}\hti_A(f,\hatn) F_a^A(\hatn) \, \,   e^{-2\pi i f(t-\hatn\cdot\vx_a /c )}
\, ,
\ees
and therefore
\be\label{tildehaFA}
\tilde{h}_a(f)=\int_{S^2} d^2\hatn \sum_{A=+,\times}\hti_A(f,\hatn) F_a^A(\hatn) \, \,   
e^{2\pi i f \hatn\cdot\vx_a /c}\, .
\ee
The functions $F_a^A(\hatn)$  are called the pattern functions of the $a$-th detector, and depend on the propagation direction $\hatn$ of the wave,
as well as on the detector's position and orientation.\footnote{They also depend on the system of axes that the observer uses to define the plus and cross polarization; this freedom can be parameterized by a polarization angle $\psi$, that we will not write explicitly here; see Sec.~7.2 of Ref.~\cite{Maggiore:2007ulw} for details.} The size of the detector
is taken here to be 
much smaller than the GW wavelength, a condition very well satisfied by ground-based interferometers (as the GW frequency $f$ ranges between 10~Hz and 1~kHz, the wavelength $c/f$ ranges between $3\times 10^4$~km and 300~km), so we can neglect the spatial variation
of the GW over the  extension of the detector (see Section~4.1 of Ref.~\cite{Lawrence:2023buo} for expressions valid also away from this ``small antenna limit'').

If the signal is given by 
a cosmological stochastic background of GWs  that is stationary,  isotropic and unpolarized, its two-point correlator  is given by
\bees
\label{ave}
&&\hspace*{-10mm}\langle \tilde{h}_A^*(f,\hatn)
\tilde{h}_{A'}(f',\hatn')\rangle_U \nn\\
&&=
\delta_T (f-f')\, \frac{\delta(\hatn-\hatn')}{4\pi}\,
 \delta_{AA'}\, \frac{1}{2}S_h(f)\, ,
\ees
where here the average is actually over realizations of the ``Universe'', and again  is relative to a finite observation time $T$ so, correspondingly, we used the regularized delta function $\delta_T (f-f')$.
In \eq{ave},  $\delta(\hatn-\hatn')$ is a Dirac delta over the two-sphere,  
\be
\delta(\hatn-\hatn')=\delta (\phi -\phi ') \delta (\cos\theta -\cos\theta ')\, ,
\ee
and  $(\theta ,\phi)$ are the polar angles that define $\hatn$. The function $S_h(f)$ is the spectral density of the stochastic background. Similarly to \eq{Snreality}, it satisfies
\be\label{Shreality}
S^*_h(f)=S_h(f)\, ,\qquad S_h(-f)=S_h(f)\, .
\ee
If, furthermore, the background is Gaussian, it is only characterized by its two-point function. 
The factor $1/(4\pi)$ in \eq{ave} is a choice of normalization such that
\bees
&&\hspace*{-20mm}\int_{S^2} d^2\hatn\, d^2\hatn'
\langle \tilde{h}_A^*(f,\hatn)
\tilde{h}_{A'}(f',\hatn')\rangle_U\nn\\
&&=
\delta_T (f-f')\delta_{AA'}
\frac{1}{2}S_h(f),\,  \label{norm}
\ees
where $d^2\hatn=d\cos\theta d\f$, so that,  for each polarization, $S_h(f)$ can be naturally compared to $S_{n,a}(f)$ in \eq{Sn1}.

For  the stochastic background generated by the superposition of astrophysical sources in an observation time $T$, we can  write
\be\label{tildehdeltan}
\tilde{h}_A(f,\hatn)=\sum_{i=1}^{\nev} \tilde{h}_{A,i}(f)\, \delta(\hatn-\hatn_i)\, ,
\ee
where $\hatn_i$ is the propagation direction of the $i$-th signal, and $\nev$ is the number of events reaching the detector  in the observation time $T$. Then, \eq{snrhab} becomes
\bees\label{snrhabdiscrete}
&&\hspace*{-14mm}h_{kl}(t,\vx)
=\sum_{i=1}^{\nev} \int_{-\infty}^{\infty} df \,\nn\\
&&\hspace*{-4mm}\times\hspace*{0mm}\sum_{A=+,\times}  \tilde{h}_{A,i}(f) e^A_{kl}(\hatn_i) \, \,   e^{-2\pi i f(t-\hatn_i\cdot\vx /c )}
\, ,
\ees
and \eq{tildehaFA} becomes
\be\label{tildeaFAastro}
\tilde{h}_a(f)=\sum_{i=1}^{\nev}\sum_{A=+,\times} \tilde{h}_{A,i}(f)  F_a^A(\hatn_i) \, \,   
e^{2\pi i f \hatn_i\cdot\vx_a /c}\, .
\ee
For an astrophysical background, assumed to be unpolarized, one can show that
\bees
\label{ave_astro}
&&\hspace*{-10mm}\langle \tilde{h}_A^*(f,\hatn)
\tilde{h}_{A'}(f',\hatn')\rangle_U \nn\\
&&=
\delta_T (f-f')\, \frac{\delta(\hatn-\hatn')}{4\pi}\,
 \delta_{AA'}\, \frac{1}{2}S^{\rm astro}_h(f)\, ,
\ees
where $S^{\rm astro}_h(f)$ is the spectral density due to the superposition of $\nev$ compact binary coalescences (CBCs)  that take place over an observation time $T$, and  is given by\footnote{Despite the fact that this expression is commonly used in the literature, we have not been able to find a convincing derivation of it, and we provide one in the companion paper Ref.~\cite{paper:spectraldensity}, where we show how it emerges from a simultaneous  average over arrival times, polarization angles, arrival directions and inclination of the orbits of the ensemble of CBCs. 
In Ref.~\cite{paper:spectraldensity}
we also show that in the correlator (\ref{ave_astro}) can also appear terms  associated to circular and even linear polarization  because of shot noise, i.e. because of the fact that in the observation time $T$ we only have  a finite number of events $\nev$, so the averages that set to zero the Stokes parameters associated to polarization are not exactly realized, and we  demonstrate how shot noise generates spatial anisotropies even when the underlying astrophysical distribution is isotropic (see also Refs.~\cite{Jenkins:2019uzp,ValbusaDallArmi:2023ydl}).}
\be\label{eq:Shsum}
S_h^{\rm astro}(f)= \frac{1}{T}
\sum_{i=1}^{\nev}
\[  |\tilde{h}_{+,i}(f)|^2 + |\tilde{h}_{\times,i}(f)|^2 \] \, .
\ee
The energy density associated to a stochastic background is related to its spectral density by~\cite{Allen:1996vm,Allen:1997ad,Maggiore:1999vm} 
\be\label{rho3}
\frac{d\rho_{\rm gw}}{d\log f}=\frac{\pi c^2}{2G}f^3S_h(f)\, .
\ee
It is convenient to define the dimensionless quantity
\be \label{eq: Omegagw general def}
\omgw(f) \equiv \frac{1}{\rho_c}\,\frac{ d\rho_{\rm gw}}{d\log f} \, ,
\ee
where  $\rho_c = 3 c^2 H_0^2/\left(8\pi G\right)$ is the critical energy density for closing the Universe. 
In terms of $S_h(f)$, we  have
\be\label{eq:OmegaSh}
\omgw(f) = \frac{4\pi^2}{3H_0^2}\, f^3S_h(f)\, .
\ee
For an astrophysical background, therefore,
\be\label{eq:Omegagw_tot_def}
\omgwa(f)=\dfrac{4\pi^2}{3H_0^2} \dfrac{f^3}{T}\sum_{i=1}^{\nev} \[ \, |\tilde{h}_{+,i}(f)|^2 + |\tilde{h}_{\times,i}(f)|^2\]\, .
\ee
Observe that, in the limit of large observation time, this expression converges to a finite quantity. This can be better seen rewriting \eq{eq:Omegagw_tot_def} as
\bees\label{eq:Omegagw_tot_rate}
\omgwa(f)&=&\dfrac{4\pi^2}{3H_0^2}\, f^3 \dfrac{\nev}{T}\\
&&\hspace*{-0mm}\times\frac{1}{\nev}\sum_{i=1}^{\nev} \[ \, |\tilde{h}_{+,i}(f)|^2 + |\tilde{h}_{\times,i}(f)|^2\]\, .\nn
\ees
In the large $T$ limit, $\nev/T$ converges to the merger rate, while the term in the second line is the average of 
$|\tilde{h}_{+,i}(f)|^2 + |\tilde{h}_{\times,i}(f)|^2$ over the population of events.

\section{The cross-correlation method for stochastic backgrounds}\label{sect:cross}

In this section we  first recall, in \cref{sect:optimal}, the  strategy for searching for stochastic GW backgrounds with a detector network by cross correlating the outputs of two detectors,   in the simplest scenario of a stochastic GW background which is  stationary, unpolarized, isotropic and Gaussian, and two detectors with uncorrelated noise. This  material  is standard~\cite{Allen:1996vm,Allen:1997ad,Maggiore:1999vm,Maggiore:2007ulw,Romano:2016dpx}, but still it will be useful to fix the notation and prepare the stage for  further extensions. In particular, in \cref{sect:correlnoise} we will discuss the case of correlated noise, which will be  important later, since we will eventually treat the residual astrophysical background (after subtraction of resolved sources, and including the errors in the subtraction) as an effective correlated noise among the detectors, in a two-detector correlations. For completeness, further extensions will be briefly mentioned in \cref{sect:further}.

\subsection{The ``vanilla'' scenario}\label{sect:optimal}

We recall here the procedure for maximizing the signal-to-noise ratio of a two-detector correlation, assuming that the noise in the two detectors are not correlated, and that the  GW background  is  stationary, unpolarized, and isotropic.

Given two GW detectors, labeled by indices $a, b$ (with $a\neq b$), taking data over a time $T$, one forms the correlator of their outputs, 
\be\label{defYab}
Y_{ab}=\int_{-T/2}^{T/2}dt\int_{-T/2}^{T/2}dt'\, s_a(t)s_b(t')Q_{ab}(t-t')\, ,
\ee
where $Q_{ab}(t-t')$ is a (real) filter function that falls rapidly to zero as $|t-t'|\ra\infty$.  If the useful bandwidth of the detector is at frequencies $f$ such that $fT\gg 1$ (which,  for 3G  ground-based detectors, operating at $f\,\gsim\, (5-10) $~Hz, is already true for a stretch of data corresponding, e.g., to $T=60$~s), the above expression can be rewritten as
\be
Y_{ab}= \int_{-\infty}^{\infty}df\, \tilde{s}^*_a(f)\tilde{s}_b(f)\tilde{Q}_{ab}(f)\, .
\ee
The filter function $Q_{ab}$ is then chosen  as follows. The quantity $Y_{ab}$ is a stochastic variable, since it depends on the noise realization and, in the presence of a cosmological background, also on the cosmological signal, which is itself a stochastic variable. One then defines $S_{ab}$ as the ensemble average of $Y_{ab}$, and $N_{ab}$ as its root-mean-square  value,
\bees
S_{ab}&=&\langle Y_{ab} \rangle\, ,\label{defSab}\\
N_{ab} &=&  \[ \langle Y^2_{ab} \rangle - \langle Y_{ab} \rangle^2 \]^{1/2} \label{defNab}\, ,
\ees
where $\langle \ldots \rangle$ denotes an ensemble average both over noise realizations and over realizations of the astrophysical or cosmological stochastic signals [realizations of the ``Universe'', see \eq{defaveragenU}].
We then look for the filter function that maximizes the signal-to-noise ratio\footnote{Note that, to avoid confusion, we reserve the notation SNR to the signal-to-noise ratio of individual CBCs in a detector network,  obtained through matched filtering to their waveform, and $(S/N)_{ab}$ to the signal-to-noise ratio of a stochastic background in a detector pair $(a,b)$, obtained through the correlation procedure that we are describing here.} 
\be\label{defSsuNab}
\(\frac{S}{N}\)_{ab}\equiv \frac{S_{ab}}{N_{ab}}\, .
\ee
We are interested in extracting a small signal from a much larger noise, so we work to lowest order in $h$. For uncorrelated noise, as we will see in a moment, the numerator $S_{ab}$ is already ${\cal O}(h^2)$ so, in $(S/N)_{ab}$, the lowest-order term is obtained setting $h=0$ in the  denominator, 
\be\label{N2ave}
N_{ab} \simeq \[ \langle Y^2_{ab} \rangle - \langle Y_{ab} \rangle^2 \]^{1/2}_{|h=0}\, .
\ee
The signal $S_{ab}$ is  then computed writing 
\be\label{SabcorrQ}
S_{ab}=\int_{-\infty}^{\infty}df\,
\langle\tilde{s}^*_a(f)\tilde{s}_b(f)\rangle \tilde{Q}_{ab}(f)\, ,
\ee
and using $\tilde{s}_a(f)= \tilde{n}_a(f)+\tilde{h}_a(f)$, and similarly for $\tilde{s}_b(f)$.
Assuming (crucially) that the noise in the two detectors are uncorrelated, and that they are also uncorrelated with the signal,  
we have
\be\label{langlesshh}
\langle\tilde{s}^*_a(f)\tilde{s}_b(f)\rangle=
\langle\tilde{h}^*_a(f)\tilde{h}_b(f)\rangle_U\, ,
\ee
where we used the fact that, on $\tilde{h}^*_a(f)\tilde{h}_b(f)$, $\langle\ldots\rangle$ reduces to 
$\langle\ldots\rangle_U$. Using \eq{tildehaFA} together with \eq{norm} (which  assumes that the background is 
stationary, isotropic and unpolarized), we get
\be\label{sscorffp}
\langle\tilde{h}^*_a(f)\tilde{h}_b(f')\rangle_U=\delta_T(f-f')\frac{1}{2} S_h(f) \Gamma_{ab}(f)\, ,
\ee
where  
\be\label{defGamma}
\Gamma_{ab}(f)=\int\frac{d^2\hatn}{4\pi}\sum_A F_a^A(\hatn)F_b^A(\hatn)\, e^{-i 2\pi f\hatn\bdot (\vx_a-\vx_b)/c}\,, 
\ee
is  called the (non-normalized) overlap reduction function of the two detectors.\footnote{It is also common to introduce  the normalized overlap reduction function $\gamma_{ab}(f)=(5/2)\Gamma_{ab}(f)$. This has the property that, for two parallel and co-located L-shaped interferometers, $\gamma_{ab}(f)=1$. This, however, is not true in general, e.g. for the interferometers at $60^{\circ}$ that make up ET in the triangular configuration, unless one changes the normalization factor. In the following, we prefer to work with $\Gamma_{ab}(f)$.
Note also that, writing explicitly the pattern functions as $F_a^A(\hatn, \psi )$, where the angle $\psi$ defines the orientation of the axis used to define the plus and cross polarizations in the plane orthogonal to $\hatn$, 
the combination  $\sum_A F_a^A(\hatn, \psi )F_b^A(\hatn , \psi)$ is independent of $\psi$, because under such rotations   $F_a^A$ transforms as a helicity-two field (see Eqs.~(7.29) and (7.30) of Ref.~\cite{Maggiore:2007ulw}), and therefore $\sum_A F_a^A(\hatn, \psi )F_b^A(\hatn , \psi)$ is invariant under $\psi\ra\psi+\psi_0$.}  
Observe  that $F_a^{+}(-\hatn)=F_a^{+}(\hatn)$ and $F_a^{\times}(-\hatn)=-F_a^{\times}(\hatn)$.\footnote{This follows from the definition $F_a^A(\hatn )=D_a^{kl}e_{kl}^A(\hatn)$, where $D_a^{kl}$ is a fixed matrix that encodes the detector geometry 
(see Eq.~(7.21) of Ref.~\cite{Maggiore:2007ulw}), together with 
$e_{ij}^+(-\hatn)=e_{ij}^+(\hatn)$ and
$e_{ij}^{\times}(-\hatn)=-e_{ij}^{\times}(\hatn)$ (see Ref.~\cite{paper:spectraldensity} for a proof). }
Therefore the overlap reduction function is real, since, in the integrand in \eq{defGamma},  the imaginary part coming from the exponential  is odd under $\hatn\ra -\hatn$, and therefore vanishes upon integration over $d^2\hatn$. The remaining part is real and also even in $f$, so,
\be\label{Gammareality}
\Gamma_{ab}^*(f)=\Gamma_{ab}(f)\, ,\qquad \Gamma_{ab}(-f)=\Gamma_{ab}(f)\, .
\ee
Then, using \eq{deltazero},
we get 
\be\label{sscor}
\langle\tilde{h}^*_a(f)\tilde{h}_b(f)\rangle_U=\frac{T}{2} S_h(f) \Gamma_{ab}(f)\, ,
\ee
and, using \eqss{SabcorrQ}{langlesshh}{sscor}, we finally obtain
\be\label{Sabnocorr}
S_{ab}=\frac{T}{2}\int_{-\infty}^{\infty}df\, S_h(f) \Gamma_{ab}(f)\tilde{Q}_{ab}(f)\, .
\ee
The calculation of $N_{ab}$, using the lowest-order term in  $h$ given by \eq{N2ave} and assuming again uncorrelated noise, is analogous and gives
\be\label{Nab2}
N^2_{ab}=\frac{T}{4}\int_{-\infty}^{\infty}df\,  |\tilde{Q}_{ab}(f)|^2\, S_{n,ab}^2(f) \, ,
\ee
where 
\be
S_{n,ab}=\[ S_{n,a}(f) S_{n,b}(f) \]^{1/2}\, ,
\ee
and $S_{n,a}(f), S_{n,b}(f)$ are the noise spectral densities of the individual detectors.
Then
\be\label{SoverN}
\(\frac{S}{N}\)_{ab}=T^{1/2}\, \frac{\int_{-\infty}^{\infty}df\, S_h(f) \Gamma_{ab}(f)\tilde{Q}_{ab}(f)}{\[ \int_{-\infty}^{\infty}df\,  |\tilde{Q}_{ab}(f)|^2\, S_{n,ab}^2(f) \]^{1/2}}\, .
\ee
We can rewrite this in a form that makes more explicit the reality of this expression; since $Q(t)$ is a real function, we have $\tilde{Q}(f)=\tilde{Q}^*(-f)$, while $S_h(f)$ and  $\Gamma_{ab}(f)$ are real and even functions, see \eqs{Shreality}{Gammareality}, so we can transform the integrals in $df$ from $[-\infty,+\infty]$ to $[0,+\infty]$, and we get
\be\label{SoverNreal}
\(\frac{S}{N}\)_{ab}=(2T)^{1/2}\, \frac{\int_{0}^{\infty}df\, S_h(f) \Gamma_{ab}(f) {\rm Re}\big[ \tilde{Q}_{ab}(f)\big] }{\[ \int_{0}^{\infty}df\,  |\tilde{Q}_{ab}(f)|^2\, S_{n,ab}^2(f) \]^{1/2}}\, .
\ee
One now defines a scalar product between two functions $\tilde{A}(f)$ and $\tilde{B}(f)$ as
\begin{equation}\label{definner}
    \inner{A}{B} = \int_{-\infty}^{+\infty} d f \ftconj{A}(f)\ft{B}(f) \Snab^2(f) \, .
\end{equation}
Then, \eq{SoverN} can be rewritten as
\be\label{Vol1_236}
\(\frac{S}{N}\)_{ab}=T^{1/2}\, \frac{\( \Gamma_{ab}S_h/S_{n,ab}^2\,\, , \,  \tilde{Q}_{ab}  \)_{ab}}{\(\tilde{Q}_{ab}\,\, , \, \tilde{Q}_{ab} \)_{ab}^{1/2}}\, .
\ee
In this form, it becomes clear that the filter function that maximizes the signal-to-noise ratio is the ``vector'' $\tilde{Q}_{ab}$ (in the infinite-dimensional space of functions)  parallel to $\Gamma_{ab}S_h/S_{n,ab}^2$, i.e.
\be\label{Qabopt}
\tilde{Q}_{ab}(f) = {\rm const.}\, \frac{\Gamma_{ab}(f)S_h(f)}{ S_{n,ab}^2(f)}\, .
\ee
The corresponding optimal signal-to-noise ratio, for  a stochastic background described by $S_h(f)$, is then
\be\label{Vol1_239}
\(\frac{S}{N}\)_{ab}= \[ 2T\int_0^{\infty}df\,  \(\frac{\Gamma_{ab}(f) S_h(f)}{S_{n,ab}(f)}\)^2 \]^{1/2}\, .
\ee
Note that the arbitrary constant in \eq{Qabopt} canceled when inserting $\tilde{Q}_{ab}(f)$ into \eq{Vol1_236}, and that the optimal filter function is actually real.

It can also be convenient to define the characteristic  amplitude of the signal~\cite{Maggiore:1999vm}, 
\be\label{defhc}
h_c(f)=\[ 2f S_h(f)\]^{1/2}\, ,
\ee
and the characteristic  amplitude of the noise,
\be\label{defhn}
\hnab(f)=\[ \frac{2f S_{n,ab}(f)}{|\Gamma_{ab}(f)|} \]^{1/2}\, \frac{1}{(2fT)^{1/4}}\, .
\ee
In terms of these quantities, \eq{Vol1_239} reads
\be\label{SNhchn}
\(\frac{S}{N}\)_{ab}=\[ \int_0^{\infty}d\log f\, 
\(\frac{h_c(f)}{\hnab(f)}\)^4 \]^{1/2}\, .
\ee
Observe that $h_c(f)$ and $\hnab(f)$ are dimensionless; $h_c(f)$ depends only on the spectral density of the signal, while in $h_n(f)$ we collect all information about the measurement process, namely the noise spectral density of the two detectors, that enter through $S_{n,ab}(f)=[ S_{n,a}(f) S_{n,b}(f) ]^{1/2}$, the overlap reduction function $\Gamma_{ab}(f)$, and the observation time $T$. Note also that $\hnab(f)$ diverges in correspondence with the zeros of the overlap reduction function.

\Eq{Vol1_239}, or equivalently \eq{SNhchn},  give the signal-to-noise ratio of a stochastic background, obtained from the correlation of two detectors. Let us recall that the assumption behind it are that  the background is stationary, unpolarized and isotropic [since we used \eq{norm}], that the noise in the two detectors is uncorrelated [which entered in \eq{langlesshh}] and that the signal in a single detector is much lower than the noise [since we used \eq{N2ave}]. 
A number of extensions of this  scenario are, however, important, and we will discuss them in the following subsections.

\subsection{Correlated noise}\label{sect:correlnoise}

The assumption that noise is not correlated between the two detectors is in general simplistic. For instance, a possible source of broad-band noise that can affect the correlation between ground-based detectors even when they are located at far away locations on the Earth are  Schumann resonances in the Earth's magnetic field. These resonances can be excited  by lightning strikes,  producing coherent oscillations
over thousands of kilometers, and have indeed been  been observed to produce correlations across the LIGO and Virgo sites~\cite{Thrane:2013npa,Thrane:2014yza,Coughlin:2016vor}.   The effect of the Schumann resonances and other forms of correlated magnetic noise on  ET has been investigated recently in Ref.~\cite{Janssens:2021cta}.
Other forms of correlated noise, such as seismic and Newtonian noise, only act  on more local scales, but can be particularly important for ET in its single-site triangular configuration, and have been investigated in several recent papers~\cite{Badaracco:2019vjq,NN_Sardinia2020,10.1785/0220200186,Bader:2022tdz,Koley:2022wpe,Janssens:2022xmo}
(see also Sec.~5.4 of Ref.~\cite{Branchesi:2023mws} for a comprehensive discussion of seismic, Newtonian and magnetic noise at ET).

For our purposes, correlated noise will  be important because the astrophysical  CBC background due to unresolved sources, as well as the error in the subtraction of resolved sources, will effectively act as a correlated noise, that could mask a cosmological background. Let us therefore see how the above results on the optimal signal-to-noise ratio are affected by correlated noise. 
Assuming that also the correlated noise component between detectors $a$ and $b$ is stationary, in analogy to \eq{Sn1} we define the correlation strain matrix ${\cal N}_{ab}$ as
\be\label{calNab}
\left\langle \tilde{n}_a^*(f)\tilde{n}_b(f) \right\rangle_n =
\delta_T(f-f')\frac{1}{2}{\cal N}_{ab}(f) \, ,
\ee
so that, when $a=b$, ${\cal N}_{aa}(f)=S_{n,a}(f)$, while the off-diagonal terms with $a\neq b$ represent the correlated component of the noise among  two detectors (and are sometimes called the cross-PSD).  Taking the complex conjugate of \eq{calNab} we find 
\be\label{calNabstar}
\left\langle \tilde{n}^*_b(f')\tilde{n}_a(f) \right\rangle_n =
\delta_T(f-f')\frac{1}{2}{\cal N}^*_{ab}(f) \, .
\ee
On the other hand, the correlator on the left-hand side is also the same as 
$\delta_T(f-f')(1/2){\cal N}_{ba}(f)$, from
which it follows that ${\cal N}_{ab}(f)$ is Hermitian,
\be\label{calNHerm}
{\cal N}^*_{ab}(f)={\cal N}_{ba}(f)\, .
\ee
Furthermore, the reality of $n_a(t)$ implies that $\tilde{n}^*_a(f)=\tilde{n}_a(-f)$, and similarly for $n_b(t)$. Replacing $f\ra -f$ and $f'\ra -f'$ in \eq{calNab}, we then obtain
${\cal N}_{ab}(-f)={\cal N}_{ba}(f)$ which, combined with \eq{calNHerm}, implies
\be\label{calNreality}
{\cal N}^*_{ab}(f)={\cal N}_{ab}(-f)\, .
\ee
From \eqs{SabcorrQ}{sscor}, the signal $S_{ab}$ (with $a\neq b$) is now given by
\bees
S_{ab}&=&\int_{-\infty}^{\infty}df\, \[
\langle\tilde{n}^*_a(f)\tilde{n}_b(f)\rangle + \langle\tilde{h}^*_a(f)\tilde{h}_b(f)\rangle 
\]
\tilde{Q}_{ab}(f)\nn\\
&=& \frac{T}{2} \, \int_{-\infty}^{\infty}df\, \[ {\cal N}_{ab}(f) + S_h(f) \Gamma_{ab}(f) \] \tilde{Q}_{ab}(f)\, ,
\label{SabcorrQcorrN}
\ees
where we made the  assumption that the instrumental noise and the cosmological signal are not correlated (neither in a given detector nor among different detectors). We write
\be
S_{ab}=S_{ab}^{\cal N}+S_{ab}^{\cal H}\, ,
\ee
where 
\bees
S_{ab}^{\cal N} &=& \frac{T}{2} \, \int_{-\infty}^{\infty}df\,  {\cal N}_{ab}(f) \tilde{Q}_{ab}(f)\, ,\\
S_{ab}^{\cal H}&=&\frac{T}{2} \, \int_{-\infty}^{\infty}df\,  
S_h(f) \Gamma_{ab}(f)  \tilde{Q}_{ab}(f)\, .
\ees
Using \eqss{Shreality}{Gammareality}{calNreality} we can rewrite these expressions in a form that makes more explicit the fact that they are real,  
\bees
S_{ab}^{\cal N} &=&  T  \int_{0}^{\infty}df\,  {\rm Re} \big[ {\cal N}_{ab}(f) \tilde{Q}_{ab}(f)\big]\, ,\label{defSabcalN}\\
S_{ab}^{\cal H} &=& T  \int_{0}^{\infty}df\,  S_h(f) \Gamma_{ab}(f) {\rm Re} \big[    \tilde{Q}_{ab}(f) \big]\, .\label{defSabcalH}
\ees
Observe that both $S_{ab}^{\cal N}$ and $S_{ab}^{\cal H}$
grow as $T$, which can eventually be traced to \eqs{ave}{calNab} with $f=f'$, together with \eq{deltazero}. Therefore, a correlated noise cannot be beaten by integrating the correlated output for a long time. This is in contrast to \eq{Vol1_239} for uncorrelated noise, where the signal-to-noise ratio  grows as $T^{1/2}$; physically, this  reflected the fact that the signal-signal correlator grows like $T$, while the noise-noise correlator, for uncorrelated noise, performs a random walk and only grows as $T^{1/2}$.
To deal with correlated noise, therefore, we must resort to different strategies.  A first handle is to try to measure it as well as possible (e.g., in the case of environmental noise, using witness sensors) and subtract it (see e.g. Ref.~\cite{Thrane:2014yza}); another option is to model its frequency dependence and, if it is different from that of a specific cosmological background that we are searching, we can exploit this feature to separate, to some extent, the two contributions  with a Bayesian analysis (similarly to what is done when separating two genuine cosmological signals, see Refs.~\cite{Smith:2017vfk,Biscoveanu:2020gds,Martinovic:2020hru}).
What is left in \eq{SabcorrQcorrN} is then the part of the correlated noise that we have been unable to subtract or characterize accurately.

Then, with the understanding that ${\cal N}_{ab}(f)$ in \eq{SabcorrQcorrN} is this ``irreducible'' part of the correlated noise that we cannot further model and subtract, a first condition for the detectability of a cosmological signal is that it stays above the correlated noise,
\be\label{SabBlargerthanSabN}
\left| \frac{S_{ab}^{\cal H}}{ S_{ab}^{\cal N}}\right| >1\, ,
\ee
i.e.
\be\label{ratiosignaltoNcorr}
\left|\frac{\int_{0}^{\infty}df\, S_h(f) \Gamma_{ab}(f)\, {\rm Re} \big[  \tilde{Q}_{ab}(f)\big]}
{\int_{0}^{\infty}df\, {\rm Re} \big[{\cal N}_{ab}(f) \tilde{Q}_{ab}(f)\big]}\right| > 1\, .
\ee
(This condition is Eq.~5.113 of Ref.~\cite{Allen:1997ad}, written in our notation). Observe that, while in the case of uncorrelated noise and optimal filter function  $(S/N)_{ab}$ is positive definite, see
\eq{Vol1_239},
for a generic filter function (and generic correlated noise) $S_{ab}^{\cal H}$ and $S_{ab}^{\cal N}$ are not necessarily positive definite. When studying the effect of the astrophysical confusion noise on a cosmological search, we will be interested in situations where we correlate the outputs of two detectors  containing the astrophysical contribution, with a filter  optimized for the cosmological searches rather than for the astrophysical background itself, so we will be interested in keeping the filter function generic. Then,
we have imposed  conditions  such as that in \eq{SabBlargerthanSabN}  on the corresponding absolute values. 
One could also require this condition to hold only in a range of frequency $[f_1,f_2]$; however, this condition can be simply inserted in the choice of filter function, requiring it to vanish outside the interval $[f_1,f_2]$. Similarly, if one is interested only in a very small range of frequencies, one can require this condition to hold just point-like, which for a real $\tilde{Q}(f)$ gives 
\be
|S_h(f) \Gamma_{ab}(f)|>
|{\rm Re}\,  {\cal N}_{ab}(f)|\, ,
\ee
which is Eq.~(5.109) of Ref.~\cite{Allen:1997ad}, in our notation.

We also define\footnote{In this definition,  it is understood that $N_{ab}$ at the denominator is computed for $h=0$. We do not write this explicitly in order not to burden the notation.}
\be\label{defSHsuN}
\(\frac{S_{\cal H}}{N}\)_{ab}\equiv \left| \frac{S_{ab}^{\cal H}}{N_{ab}}\right| \, ,
\ee
and 
\be\label{defSNsuN}
\(\frac{S_{\cal N}}{N}\)_{ab}\equiv \left|\frac{S_{ab}^{\cal N}}{N_{ab}}\right| \, ,
\ee
so, in particular, $(S_{\cal N}/N)_{ab}$ is a signal-to-noise ratio in which the ``signal'' is actually the correlated noise, and the time integration extracts it from the much larger uncorrelated noise described  by $S_{n,ab}(f)$.\footnote{We are assuming here that the part of noise which is correlated among the two detectors  is a small fraction of the total noise of each detector, so that, at the denominator of \eqs{defSHsuN}{defSNsuN} we still use the quantity $N_{ab}$ computed for uncorrelated noise. See Ref.~\cite{Enis_in_prep} for the  extension to generic correlated noise.}
Then, \eq{SabBlargerthanSabN} can also be expressed in the form
\be\label{newfloor}
\(\frac{S_{\cal H}}{N}\)_{ab} > \(\frac{S_{\cal N}}{N}\)_{ab}\, ,
\ee
or, more generally, $(S_{\cal H}/N)_{ab}> \kappa\, (S_{\cal N}/N)_{ab}$ where $\kappa$ depends on the confidence level required for detection (e.g., $\kappa=1$, as above, or $\kappa=3$, to mention another value typically used).

Furthermore, of course, the contribution  of the cosmological background to the signal-to-noise ratio must still be above one (or whatever threshold level we wish to impose), i.e.
\be\label{SoverN3}
\(\frac{S_{\cal H}}{N}\)_{ab} > 1\, .
\ee
Therefore, a cosmological background is now detectable if, for a suitable choice of the filter function, it satisfies both \eqs{newfloor}{SoverN3}, i.e.
\be\label{SoverNmax}
\(\frac{S_{\cal H}}{N}\)_{ab} > {\rm max}\left[ 1, \(\frac{S_{\cal N}}{N}\)_{ab} \right]\, ,
\ee
or, equivalently,
\be
\(\frac{S_{\cal H}}{N}\)_{ab} >1 \,\quad  {\rm and}\quad 
\(\frac{S_{\cal H}}{S_{\cal N}}\)_{ab}>1\, .
\ee
Explicitly, from \eqs{Nab2}{defSabcalH}
\be\label{SoverN2}
\(\frac{S_{\cal H}}{N}\)_{ab}=(2T)^{1/2}\, \frac{\left| \int_{0}^{\infty}df\, S_h(f) \Gamma_{ab}(f){\rm Re} \big[  \tilde{Q}_{ab}(f)\big]\right| }{\[ \int_{0}^{\infty}df\,  |\tilde{Q}_{ab}(f)|^2\, S_{n,ab}^2(f) \]^{1/2}}\, ,
\ee
while, from \eqs{Nab2}{defSabcalN},
\be\label{SoverNdelnoise}
\(\frac{S_{\cal N}}{N}\)_{ab}=(2T)^{1/2}\, \frac{\left| \int_{0}^{\infty}df\, {\rm Re} \big[{\cal N}_{ab}(f) \tilde{Q}_{ab}(f)\big]\right|}{\[ \int_{0}^{\infty}df\,  |\tilde{Q}_{ab}(f)|^2\, S_{n,ab}^2(f) \]^{1/2}}\, .
\ee
The next question is the choice of the optimal filter function $\tilde{Q}_{ab}(f)$.
In typical situations, correlated noise is much smaller than the noise spectral densities in each detector, i.e.
\be\label{N12small}
|{\cal N}_{12}| \ll {\cal N}_{11}\, ,\qquad |{\cal N}_{12}| \ll  {\cal N}_{22} \, .
\ee
We will see that this indeed the case when the correlated noise is given by the astrophysical background. When this is the case, the crucial maximization procedure will be the one of the ratio in \eq{SoverN2}, 
and the optimal filter function will be  basically the same as in \eq{Qabopt}.

\subsection{Further extensions}\label{sect:further}

To conclude this section we briefly mention, for completeness, some further extensions of the simplest scenario where the GW background is assumed to be stationary, unpolarized, isotropic and Gaussian.

A first significant extension is to 
non-Gaussian stochastic backgrounds.
The maximization  of the signal-to-noise ratio obtained from the two-point correlator, discussed in the previous subsections, is valid independently of the
assumption that the stochastic background is Gaussian.  For a Gaussian field, however, the two-point correlator contains all the statistical information and the higher-point correlators are derived from it; therefore, no further gain is possible by considering higher-order correlator, and
\eq{Vol1_239}  gives the optimal signal-to-noise ratio.
This is not necessarily true for non-Gaussian stochastic fields, for which other strategies, involving higher-order correlators, could give better results~\cite{Drasco:2002yd,Thrane:2013kb,Martellini:2014xia,Martellini:2015mfr,Romano:2016dpx,Smith:2017vfk,Ballelli:2022bli,Renzini:2022alw,Lawrence:2023buo}, so the optimal signal-to-noise ratio could in principle  be higher compared to that obtained maximizing the two-point correlator.

In general, a background is Gaussian if, at any given moment of time, it is produced by the superposition of a large number of signals, so the relevant parameter is the ratio between the average duration of a signal (or, rather, of the part of it that gives a non-negligible contribution to the detector output) and the average time separation of the signals. More precisely, in a time-frequency plane, the background will be Gaussian in the frequency bins that, at any given time, receive contributions from a large number of signals.

From this point of view, despite the fact that (as we will see below) the BBH and BNS backgrounds have comparable energy densities, their statistical properties   are very different. The BBH background is highly non-Gaussian; in the bandwidth above 10~Hz, the average duration of the waveform, at current 2G detectors, is $0.2-0.7$~s~\cite{Regimbau:2016ike}, while the average time separation is of order of a few minutes; therefore the chances of having even just two overlapping signals are quite small.  Even at 3G detectors the chances of overlapping BHs signal are small, as we  will discuss below.
Lighter sources such as BNSs, in contrast, can stay in the band much longer, and  at 3G detectors BNSs will stay in the bandwidth for hours or up to a day. So,  the BNS background is expected to be quasi-Gaussian in most of the frequency bins. Note, however, that the BNS merger phase, which  produces the high-frequency part of the signal, is very short and most of the BNS mergers will still be quite separated in time, so even for BNS the non-Gaussianity can be quite significant in the high-frequency bins. In this paper, we will study how to subtract
the astrophysical background from the two-point correlator.
It should, however, be kept in mind that, especially for BBHs, or for BNS in the high-frequency regime, further improvements exploiting the non-Gaussianity of the astrophysical signal could in principle be possible.

Another significant extension is to multiple detectors.
When we have a network  with more than two detectors, for stochastic backgrounds the simplest strategy is to consider separately all possible detector pairs. With $N_d$  independent detectors, we can form the two-detector correlation for each of the   $N_d(N_d-1)/2$  detector pairs,  and compute the corresponding value of $S/N$. The total  signal-to-noise ratio of the detector network is then obtained by summing up in quadrature the $S/N$ of the various detectors pairs (see e.g. Ref.~\cite{Maggiore:2007ulw}, page~409). Other possibilities involving multiple detector correlations are however also possible, see Refs.~\cite{Allen:1997ad,Zhu:2011bd}. In this paper, we will limit ourselves to the correlation between detector pairs.

\section{The astrophysical background as a correlated noise}\label{sect:Astrocorrelated noise}

In the literature, there have been several attempts at characterizing the effect of unresolved astrophysical sources and the error induced  by the subtraction of resolved  sources,  in terms of effective strain sensitivities $S_{\rm unres}(f)$ and $S_{\rm err}(f)$, or, equivalently,   in terms of  effective  normalized  energy densities
$\Omega_{\rm unres}(f)=(4\pi^2/3H_0^2) f^3 S_{\rm unres}(f)$ and 
$\Omega_{\rm err}(f)=(4\pi^2/3H_0^2) f^3 S_{\rm err}(f)$,
where $S_{\rm unres}(f)$ is the contribution to the spectral density from unresolved sources and $S_{\rm err}(f)$ is assumed to describe the contribution from the error from the subtraction of the resolved source; in particular, different expressions for 
$S_{\rm err}(f)$ [or, equivalently, for $\Omega_{\rm err}(f)$]  have been proposed,   involving the difference between the true and the reconstructed signals~\cite{Sachdev:2020bkk,Zhou:2022otw,Zhou:2022nmt,Pan:2023naq}.
However, as we will discuss in more detail in \cref{sect:comparison}, all the proposed expressions for $\Omega_{\rm err}(f)$ are in fact heuristic, and do not follow from any first-principle computation. 

In this section, we show how  to evaluate  
from first principles the effect of the astrophysical background in the search for cosmological backgrounds at a detector network, by performing explicitly the correlation between the outputs of pairs of detectors, from which a reconstruction (with associated error) of the resolved sources has been subtracted. The procedure will be performed taking into account the noise realization, and will then be repeated   for a large number of noise realizations.

\subsection{General strategy}\label{sect:strategy}

Let $n_a(t)$, with $-T/2 < t < T/2$, be the specific noise realization in a given observing run of duration  $T$, in the $a$-th detector. The total output of the $a$-th detector, including the instrumental noise, the astrophysical signals, and a possible cosmological signal,  is
\be
s_a(t)=n_a(t)+ \sum_{i=1}^{\nev} h^{\rm true}_{a,i}(t) + h_a^{\rm cosmo}(t)  \, ,
\ee
where $h^{\rm true}_{a,i}(t)$  is the projection of the  true $i$-th astrophysical signal onto the $a$-th detector, 
and $h_a^{\rm cosmo}(t)$ is the  cosmological  signal in the $a$-th detector. In this section we add a label ``true'' to the actual astrophysical signals, that were simply denoted by $h$ (e.g. $h_a$, $h_A$, etc.)
in the previous section, because we need to distinguish between the true values and the reconstructed ones.

We assume that an astrophysical signal is detected if the signal-to-noise ratio (SNR) of the full detector network is above a threshold value $\snrth$.\footnote{More accurate detection criteria are based on bounds on the false alarm rate. This, however, requires a detailed knowledge of the non-Gaussian noise in the detectors, and cannot be used to perform forecasts for future detectors.}
We denote by $h^{\rm obs}_{a,i}(t)$  the projection of the  reconstructed (i.e., ``observed'')  $i$-th astrophysical signal onto the $a$-th detector. The reconstruction of the injected signals can be performed at different levels of approximation. In \cref{sect:sourcepar_reconstruction} we discuss the procedure that we will use in this paper, which goes beyond the Fisher matrix approach (despite the fact that we deal with large populations) and also takes into account the specific noise realizations in the detector.

Denoting by $\sigma_a(t)$ the output of the $a$-th detector after subtraction of the reconstructed signals, we then have
\be\label{sigmaahconf}
\sigma_a(t)=n_a(t)+h_a^{\rm  conf}(t)+h_a^{\rm cosmo}(t)\, ,
\ee
where
\be\label{defhconf}
h_a^{\rm  conf}(t)=\sum_{i=1}^{\nev}\[  h^{\rm true}_{a,i}(t)- h^{\rm obs}_{a,i}(t) \theta(\snrobs_i - \snrth) \]\, ,
\ee
and $\theta(\snrobs_i - \snrth)$ is a Heaviside theta function,
that expresses the fact that an event is subtracted only if its SNR is above a given threshold $\snrth$.   The notation $h_a^{\rm  conf}(t)$  in \eq{defhconf} is meant to stress that this quantity is a ``confusion noise'', due to the unresolved sources (the terms in the sum where the theta function is zero), and to the error in the subtraction of resolved sources (the terms in the sum where the theta function is one).
In \eq{defhconf}, $\snrobs_i$ is the SNR of the $i$-th event in the given realization of the noise, which is what really determines the actual detectability of the signal. Note also that the reconstruction of the signal is obtained using the information from the whole detector network. Once we have reconstructed in this way an astrophysical signal, we can go back to the output of each individual detector of the network, and subtract from it the effect of the astrophysical signal, which is itself obtained by projecting the reconstructed GW amplitude onto the given detector, as in \eq{hFhF}.  
In Fourier space, 
\be
\tilde{\sigma}_a(f)=\tilde{n}_a(f)+  \tilde{h}^{\rm conf}_{a}(f) + h_a^{\rm cosmo}(f)  \, ,
\ee
where
\be\label{defhconfFourier}
\tilde{h}_a^{\rm  conf}(f)\hspace*{-0.7mm}=\hspace*{-0.7mm}
\sum_{i=1}^{\nev}\[  \tilde{h}^{\rm true}_{a,i}(f)- \tilde{h}^{\rm obs}_{a,i}(f) \theta(\snrobs_i - \snrth)\] \hspace*{-0.7mm}.
\ee
From the point of view of the search for a cosmological background, $\tilde{h}^{\rm conf}_{a}(f)$ is just a confusion noise that adds up to the instrumental noise.
The problem that we wish to address is therefore how to extract the cosmological signal, from the effective noise 
\be\label{defneff}
\tilde{n}^{\rm eff}_a(f)\equiv 
\tilde{n}_a(f)+  \tilde{h}^{\rm conf}_{a}(f)\, ,
\ee
using the correlation technique between detector pairs. We will assume that the instrumental noise are uncorrelated among detectors (see \cref{sect:correlnoise} for a discussion of the limitations of this approximation). However, in any case $\tilde{h}^{\rm conf}_{a}(f)$ is  correlated among the detectors, because it depends on the astrophysical signals. We will then treat it as an effective correlated noise. 

\begin{figure}[tb]
    \centering
    \includegraphics[width=1.\linewidth]{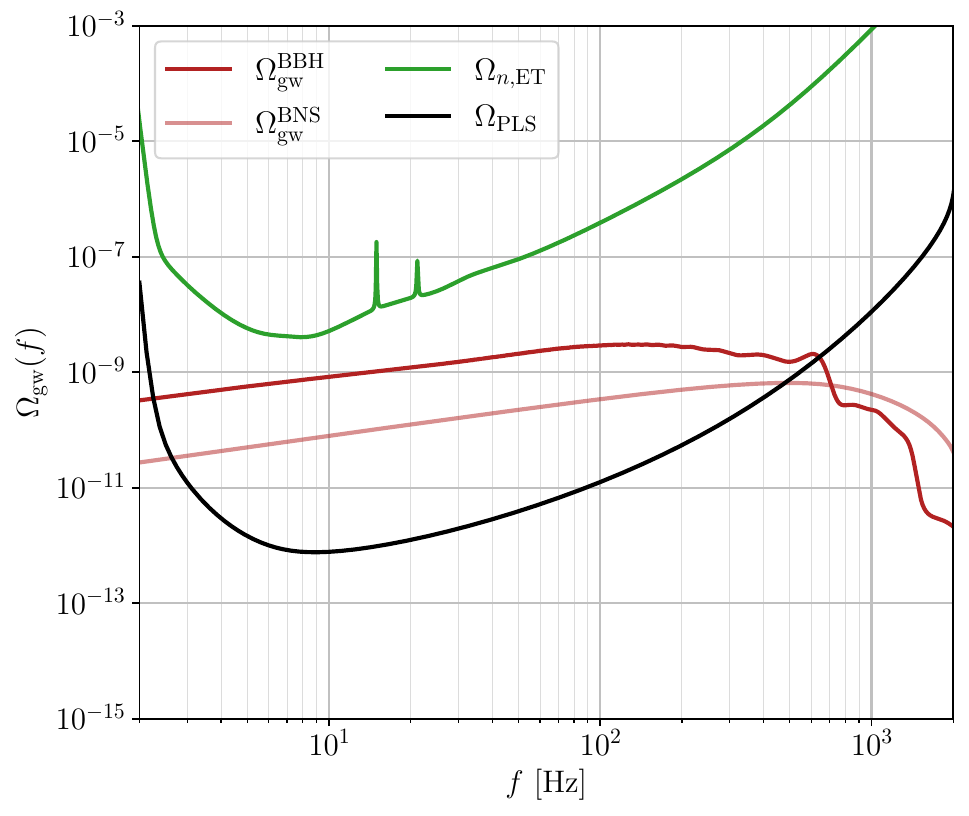}
    \caption{Comparison between the energy density $\omgw(f)$ produced by BBHs and BNSs, the equivalent energy density $\Omega_{n, \rm ET}(f) $ of the noise of a single ET detector, and the PLS curve obtained correlating for 1~yr the three ET detectors in the triangle configuration (colors as in legend).}
    \label{fig:noise_power}
\end{figure}

To understand the relative importance of $\tilde{h}^{\rm conf}_{a}(f)$ and
$\tilde{n}_a(f)$ at 3G detectors, in \cref{fig:noise_power} we show  the  energy density per unit logarithmic interval of frequency (normalized, as usual, to the critical density $\rho_c$) due to BBHs,  $\Omega_{\rm gw}^{\rm BBH}(f)$,  and that due to BNS, $\Omega_{\rm gw}^{\rm BNS}(f)$, computed using \eq{eq:Omegagw_tot_def} with our realistic population models (which is the same used and described in Ref.~\cite{Branchesi:2023mws}), before any subtraction of resolved sources. We compute it using the mergers that take place in 1~yr; given the large merger rates of BBHs and BNSs, which in our model are  $\sim1\times10^5$~events/yr and $\sim7\times10^5$~events/yr, respectively, the asymptotic limit (\ref{eq:Omegagw_tot_rate}) is reached to a good approximation, and the dependence on the sample of mergers chosen is  small, as we have indeed checked numerically comparing with samples corresponding to smaller observation times.\footnote{Some residual sample dependence appear for BBHs in the kHz region, as revealed also by the fact that the BBH curve in that region, in \cref{fig:noise_power}, is less smooth.  This is due to the fact that most BBHs merge at lower frequencies; therefore, the size of the BBH sample effectively contributing to the result in the kHz region is smaller. For the same reason, the shape of the BBH spectral density deviates earlier, compared to BNS, from the $f^{2/3}$ behavior obtained summing the CBC contributions from the inspiral phase (see also e.g. App.~A of Ref.~\cite{Zhong:2024dss}).} 
We compare them with the quantity 
\be\label{defOmeganoise}
\Omega_{n, \rm ET}(f) \equiv 
\frac{4\pi^2}{3H_0^2} f^3 S_{n, \rm ET}(f)\, ,
\ee
where $S_{n, \rm ET}(f)$  is the noise PSD of a single ET detector (out of the three that compose the ET observatory),\footnote{For ET in its triangular configuration, which is made by several nested instruments, there is some ambiguity in nomenclature. We use  the recommended terminology from the 2020 ET Design Report Update,
\url{https://apps.et-gw.eu/tds/?r=18715}: the high-frequency (HF) and low-frequency  (LF) interferometers that make the so-called ``xylophone'' configuration are indeed referred to as ``interferometers''. The combination of a HF interferometer  and a LF interferometer (whether in a L-shaped geometry, or with arms at $60^{\circ}$ as in the triangle configuration)
is called a ``detector''. Therefore ET, in its triangular geometry, is made of three nested detectors, for a total of 6 interferometers, while in its ``2L'' geometry is made of two well-separated L-shaped detectors, each with two interferometers,  for a total of 4 interferometers. The whole ensemble of detectors is called an ``observatory''.} that we have transformed into an equivalent energy fraction using \eq{eq:OmegaSh}. Let us stress that, of course, there is no actual ``energy'' associated to the PSD of a detector. The well-defined concepts are the spectral density of the noise, \eq{Sn1}, and the spectral density of the GW signal, \eq{ave}. For the GW signal, the spectral density is associated to an actual energy density through \eqs{rho3}{eq:OmegaSh}, and it can therefore be convenient to define a fictitious energy density (normalized to $\rho_c$) associated to the noise curve through \eq{defOmeganoise}. Equivalently, the comparison could be performed directly at the level of  the spectral densities 
$S_h(f)$ and
$S_{n, \rm ET}(f)$.  Transforming spectral densities into equivalent energy fractions through equations such as (\ref{defOmeganoise}) can be convenient because the latter are dimensionless quantities, while spectral densities have dimensions ${\rm Hz}^{-1}$ and in the GW context, typically unnatural values such as $10^{-44}\, {\rm Hz}^{-1}$. However, while formally any spectral density can be multiplied by the factor $4\pi^2f^3/(3H_0^2)$ to transform it into a dimensionless quantity, one should keep in mind that only for the GW spectral densities this corresponds to an actual energy density (normalized to $\rho_c$).\footnote{Similar consideration hold for the spectral densities  describing linear and circular polarization of a polarized stochastic GW backgrounds. One can formally define an equivalent energy fraction multiplying them by $4\pi^2f^3/(3H_0^2)$, but actually no real energy is associated to them; see Sec.~IV of Ref.~\cite{paper:spectraldensity} for discussion.}

The single-detector sensitivity, however, is not yet the quantity relevant for the detectability of a stochastic background, in a search performed by correlating two (or more) detectors. A useful graphical representation of the detectability of a stochastic background is given by the Power-Law integrated  Sensitivity curve (PLS), which is constructed~\cite{Thrane:2013oya}  considering power-law spectra of the form $\omgw(f)=\Omega_{\alpha} (f/f_*)^{\alpha}$, where $f_*$ is a reference frequency that can be chosen arbitrarily (its redefinition corresponds to  a redefinition of $\Omega_{\alpha}$); then, for each $\alpha$, one determines $\Omega_{\alpha}$ requiring that, performing the correlation between two detectors (or between more detectors, with the signal-to-noise ratio of the independent pairs summed in quadrature) this spectrum is detectable at $S/N=1$, with an integration time of 1~yr. The PLS is then defined as the envelope of the family of spectra obtained in this way, as $\alpha$ is varied, in principle between $-\infty$ and $+\infty$. By construction, 
every power-law spectrum that is tangent to the PLS at some frequency is detectable  with a detection threshold $(S/N)_{\rm th}=1$,
so that any power-law spectrum that, in some frequency range, is above the PLS, has 
$S/N >1$.
In \cref{fig:noise_power}
we also show  the PLS obtained by correlating the three ET detectors in the triangle configuration (for which we use the same sensitivity curves as in Ref.~\cite{Branchesi:2023mws})
with an integration time of 1~yr.

The following conclusions can be drawn from \cref{fig:noise_power}:

\begin{enumerate}

\item \label{itemone:fig1comment} Even before removing the resolved sources, the normalized energy densities $\Omega_{\rm gw}^{\rm BBH}(f)$ and $\Omega_{\rm gw}^{\rm BNS}(f)$ of the BBHs and of the BNSs  mergers is well below the equivalent energy density of the noise in a single  detector, even for  3G detectors; or, more properly, their spectral densities $S^{\rm BBH}_h(f)$ and $S^{\rm BNS}_h(f)$ are well below the spectral density of the instrumental noise of an individual 3G detector, such as $S_{n,\rm ET}(f)$. Removing the resolved sources will of course lower the residual confusion noise further (as we will compute explicitly below).
This means that, when we treat the residual astrophysical background (after subtraction of resolved sources) as a  contribution to the noise correlated among the detectors, we will be in the situation in which this correlated component of the effective noise is very small compared to the total noise in a single detector, i.e. we will be in the regime of \eq{N12small}. In other words, despite the fact that at 3G detectors there are CBC signals all the time, we see from \cref{fig:noise_power}
that their spectral density  is still a small fraction of that of the instrumental noise. 
This can be  understood physically as follows. For BBHs, the average separation of the arrival times is of order of a few minutes; the time that they remain in the bandwidth  of  3G detectors starts to be comparable, but for most events is still  smaller; e.g., a $(30+30)~\msun$ BBH  at $z\ll 1$  stays in the bandwidth for about 40~s. Furthermore,  for  given source-frame masses, the time spent in the detector's bandwidth  further decreases as we increase  the redshift:
the dependence on the redshift enters because the time to coalescence actually depends on the redshifted chirp mass ${\cal M}_c=(1+z)M_c$, see Eq.~(4.195) of Ref.~\cite{Maggiore:2007ulw}. For instance, a $(30+30)~\msun$ BBH  at $z=2$  stays in the bandwidth only for  about 6~s.
Therefore, BBH signals  are  present only a fraction of the time, while noise is always there. This is shown also in the upper panel of \cref{fig:src_strain_sim}, where we show simulated time-domain strain for 15~min of data-taking at one ET interferometer,  adding mock BBH  signals  from the populations adopted in \cref{sect:rescorr}, and we see that there are several data segments (highlighted in black) where no BBH signal is present.
Furthermore, even when a BBH is present, noise is in general much higher, and the BBH can be extracted from it only thanks to a matched filtering tuned to its waveform. This is clearly seen again from the example simulated data-stream shown in the upper panel of \cref{fig:src_strain_sim}. For BNSs the situation is more complex because, due to their smaller mass,  their signals are longer but at the same time weaker (and also their merger rate is different, and in our model somewhat larger than for BBHs). This can again be appreciated from the lower panel of \cref{fig:src_strain_sim}, in which the average event amplitude is smaller as compared to the upper panel, but a signal is always present. It should also be noticed that, while in the upper panel only 6 BBHs inspiralling and merging in the selected time window are present, in the lower we have the contribution of more than 50 BNSs in different merger phases. We see from \cref{fig:noise_power} that the net result of these effects is that (before source subtraction) the BNS contribution to the noise stays even below the one from BBHs.

\begin{figure}[tb]
    \centering
    \includegraphics[width=1\linewidth]{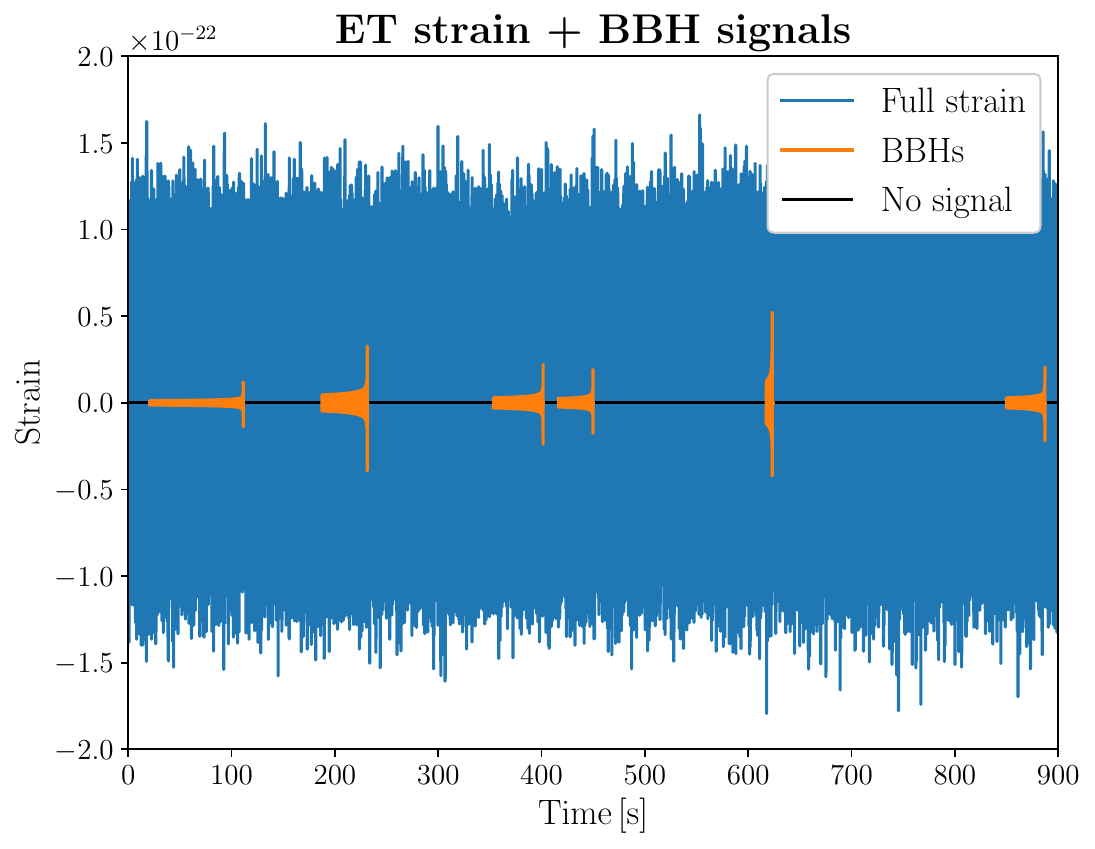}\\[.1mm]
    \includegraphics[width=1\linewidth]{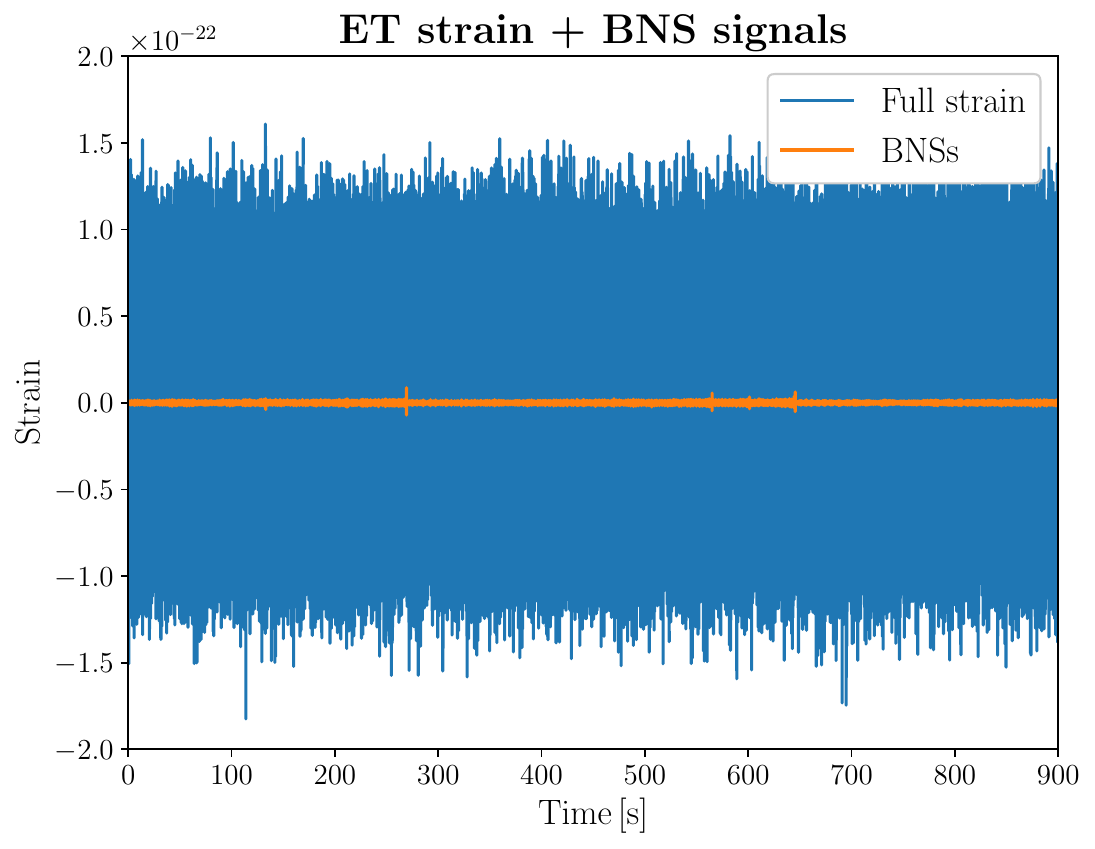}
    \caption{Representative simulated time-domain strain for 15~min of data-taking at one ET interferometer. We add mock BBH and BNS signals to the upper and lower panel, respectively, both obtained from the populations adopted in \cref{sect:rescorr}. To simulate the time-domain signals we employ the \texttt{IMRPhenomTHM} and \texttt{IMRPhenomT} approximants~\cite{Estelles:2020osj} for BBHs and BNSs, respectively. In the upper panel we highlight in black the data segments where no signal is present.}
    \label{fig:src_strain_sim}
\end{figure}

\item The PLS represents the level at which (for a power-law spectrum), using the correlation between two (or more) detectors and integrating for a long time, we can beat uncorrelated noise, pushing down the sensitivity limit. We see from \cref{fig:noise_power} that, with 1~yr of integration,  we can push down the sensitivity of ET in the triangle configuration by $2-2.5$ orders of magnitude, everywhere in frequency, compared to the single-detector sensitivity. Note that  $\Omega_{\rm gw}^{\rm BBH}(f)$ and $\Omega_{\rm gw}^{\rm BNS}(f)$ stay well above the PLS, at least up to a few hundreds of Hz.
However, this does not mean yet that the BBH and BNS  backgrounds will degrade the sensitivity to cosmological backgrounds. 
First of all, we have to remove  the resolved sources (taking, however, into account the  error in the subtraction). Furthermore, 
it is crucial to realize that the meaning of the PLS, as defined in Ref.~\cite{Thrane:2013oya}, is that a (power-law) stochastic background is detectable when it is above the PLS, {\em but only under the assumption that the two-detector correlation has been performed with the filter function that maximizes the signal-to-noise ratio of that stochastic background}.
When we search for a  given cosmological background, we will use a filter function  chosen so to maximize  its signal-to-noise ratio. This filter function is not the same  as the filter function that maximizes the search for the BBH and BNS residual stochastic backgrounds (and, as we will see, it is in fact very different). Therefore, by itself, the fact that $\Omega_{\rm gw}^{\rm BBH}$ and $\Omega_{\rm gw}^{\rm BNS}$ are above the PLS (possibly even after the subtraction of resolved sources) still tells us nothing about whether this will affect the search for a cosmological background. To understand whether this astrophysical confusion noise really has an impact, we must correlate the output of two detectors, which include this astrophysical confusion noise, {\em using the same filter function as that of the cosmological search that we want to perform} (rather than  the filter that maximizes the effect of this astrophysical confusion noise), and only then we can assess the real relevance of the  astrophysical confusion noise.

\end{enumerate}

\noindent
Our strategy for evaluating the effect of the residual astrophysical background, after subtraction of the resolved sources, will then be as follows.
Starting from the effective noise $\tilde{n}^{\rm eff}_a(f)$ defined in \eq{defneff} we introduce,  similarly to  \eq{calNab}, the function
${\cal N}^{\rm eff}_{ab}(f)$ from
\be\label{calNab1}
\left\langle \tilde{n}^{{\rm eff},*}_a(f)\, \tilde{n}^{\rm eff}_b(f') \right\rangle_n =
\delta_T(f-f')\frac{1}{2}{\cal N}^{\rm eff}_{ab}(f) \, ,
\ee
so that, setting $f'=f$, 
\be\label{calNab2}
{\cal N}^{\rm eff}_{ab}(f)=\frac{2}{T}\left\langle \tilde{n}^{{\rm eff},*}_a(f)\tilde{n}^{\rm eff}_b(f) \right\rangle_n  \, .
\ee
We now consider the signal-to-noise ratio in \eq{SoverNdelnoise} (i.e., the signal-to-noise ratio of the correlated noise with respect to the uncorrelated noise), with ${\cal N}_{ab}(f)$  given by
${\cal N}^{\rm eff}_{ab}(f)$, 
\bees\label{SoverNdelnoiseeff}
\(\frac{S_{\cal N}}{N}\)_{ab}[Q]&=&(2T)^{1/2}\, \frac{\left|\int_{0}^{\infty}df\, {\rm Re} \big[{\cal N}^{\rm eff}_{ab}(f) \tilde{Q}_{ab}(f)\big]\right|}{\[ \int_{0}^{\infty}df\,  |\tilde{Q}_{ab}(f)|^2\, S_{n,ab}^2(f) \]^{1/2}}\nn\\
&&\hspace*{-20mm}=\frac{2\sqrt{2}}{\sqrt{T}}\, \frac{\left| \int_{0}^{\infty}df\, {\rm Re} \big[
\left\langle \tilde{n}^{{\rm eff},*}_a(f)\tilde{n}^{\rm eff}_b(f) \right\rangle_n \tilde{Q}_{ab}(f)\big]\right| }{\[ \int_{0}^{\infty}df\,  |\tilde{Q}_{ab}(f)|^2\, S_{n,ab}^2(f) \]^{1/2}}
\, .\label{SoverNofQ}
\ees
We have emphasised that  $(S_{\cal N}/N)_{ab}$ is a functional of the filter function $\tilde{Q}(f)$. When searching for a specific cosmological signal,
the filter function that enters in \eq{SabcorrQcorrN}, and therefore in \eq{SoverNdelnoiseeff},
is the one that maximizes the signal-to-noise ratio of  that cosmological signal, given by \eq{Qabopt}. In a search for generic cosmological backgrounds one will scan over a family of filters and, in principle,  for each filter tested,  the signal-to-noise ratio of a cosmological signal should  be compared with \eq{SoverNdelnoiseeff}, computed with the same filter. From this we see that the level of the residuals left by the subtraction of the resolved astrophysical sources is not only a property of the detector network (i.e., how well the sources are being resolved and subtracted), but in fact also depends  on the specific cosmological search that is being performed.

According to the discussion in \cref{sect:correlnoise}, if, for a  choice  $Q(f)=\bar{Q}(f)$ that maximizes a given cosmological signal,  we find that $(S_{\cal N}/N)_{ab}[\bar{Q}]<1$,  then in \eq{SoverNmax} the dominant condition is 
$(S_{\cal H}/N)_{ab}> 1$, as in the absence of correlated noise,  so the
astrophysical confusion noise (due to the unresolved sources and to the error in the subtraction of the resolved sources)  will not affect the search for that cosmological background. If, in contrast, $(S_{\cal N}/N)_{ab}[\bar{Q}]>1$, this becomes the dominant condition in \eq{SoverNmax}; this does not mean, yet, that this will define a new floor for the sensitivity to that cosmological background. Indeed, we will see below that we will be able to characterize in any case the frequency dependence of this residual astrophysical contribution, so that, in the search for a stochastic background with a sensibly different frequency dependence, we could still be able to apply Bayesian separation techniques. However, in the search for generic stochastic backgrounds, with generic frequency dependence, this will be more difficult. So, the bottom-line is that, if $(S_{\cal N}/N)_{ab}[\bar{Q}]<1$, we can forget about the residual astrophysical background in the given cosmological search while, if $(S_{\cal N}/N)_{ab}[\bar{Q}]>1$, we might still be able to dig under it, but this will require further work, and might depend on the specific search performed.

\subsection{Evaluation of \texorpdfstring{$\nab$}{the correlator}}\label{sect:evalcorr}

The next step is to evaluate the correlator $\nab$ that appears in \eq{SoverNofQ}. Using the definitions (\ref{defhconf}) and (\ref{defneff}), we have
\be
\tilde{n}^{\rm eff}_a(f)=\tilde{n}_a(f)+ 
\sum_{i=1}^{\nev}[ \tilde{h}^{\rm true}_{a,i}(f)- \tilde{h}^{\rm obs}_{a,i}(f) \theta_i] \, ,
\ee
where, to have a more compact notation, we have introduced
\be\label{defthetai}
\theta_i\equiv \theta(\snrobs_i - \snrth)\, ,
\ee
(observe that $\snrobs_i$ is the SNR of the $i$-th event in the whole detector network, and therefore does not carry a detector index $a$).
Then
\bees\label{eq: correlator_subtraction_full1}    
\neff &=&
\big\langle
\big\{
\tilde{n}^*_a+ 
\sum_{i=1}^{\nev} (\tilde{h}^{\rm true}_{a,i}- \tilde{h}^{\rm obs}_{a,i} \theta_i)^* \big\}\nn\\
&&\hspace{-2mm}\times
\big\{\tilde{n}_b+ 
\sum_{j=1}^{\nev}(\tilde{h}^{\rm true}_{b,j}- \tilde{h}^{\rm obs}_{b,j} \theta_j) 
\big\}
\big\rangle\, ,
\ees
where, again for notational simplicity,  it is now understood that all Fourier modes are evaluated at the same frequency $f$. 
We are interested in computing this quantity for $a\neq b$, and we are assuming that the instrumental noise among detectors are uncorrelated, so  
$\langle\tilde{n}^*_a\tilde{n}_b\rangle_n=0$. Observe, however, that the cross term between the instrumental noise $\tilde{n}^*_a$ and $(\tilde{h}^{\rm true}_{b,i}- \tilde{h}^{\rm obs}_{b,i} \theta_i)$ is non-zero, because the noise realization in all detectors of the network affects the reconstruction of the signal, so both $\tilde{h}^{\rm obs}_{b,i}$ and $\theta_i$ depend on the noise in all detectors.
In contrast, $\tilde{h}^{\rm true}_{a,i}$, being the true signal, is independent of the noise, and can be carried out from the average over noise realizations; nevertheless, $\tilde{h}^{\rm true}_{a,i}$ is still a stochastic variable because it  depends on the ``Universe'' realization, i.e., in this case, on the realization of the discrete stochastic astrophysical signal. We use  \eq{defaveragenU}
and compute first the average over the noise realizations, for a fixed Universe realization.
Then, using also the fact that $\langle n_a\rangle_n=0$, we get
\bees
&&{\neff}_n =
\sum_{i,j=1}^{\nev} \[ 
\tilde{h}^{{\rm true},*}_{a,i}\tilde{h}^{\rm true}_{b,j}
+\langle \tilde{h}^{{\rm obs},*}_{a,i}\theta_i\tilde{h}^{\rm obs}_{b,j} \theta_j\rangle_n 
 \right. \nn\\
&&\hspace*{20mm}\left.
-\langle \tilde{h}^{{\rm obs}, *}_{a,i} \theta_i\rangle_n \tilde{h}^{\rm true}_{b,j}
-\tilde{h}^{{\rm true},*}_{a,i}\langle \tilde{h}^{\rm obs}_{b,j} \theta_j\rangle_n
\]\nn\\
&&\hspace*{20mm}-\sum_{i=1}^{\nev} \langle \theta_i (\tilde{n}^*_a\tilde{h}^{\rm obs}_{b,i}+ \tilde{n}_b\tilde{h}^{{\rm obs},*}_{a,i} )\rangle_n
\, .\label{eq: correlator_subtraction_full2} 
\ees
(Note  that the sum over events commutes with the average over noise realizations).
This can also be rewritten as
\bees\label{eq: correlator_subtraction_full3} 
&&\hspace*{-6mm}{\neff}_n =\nn\\
&&\hspace*{-2mm}\sum_{i=1}^{\nev}  \(\tilde{h}^{\rm true}_{a,i}- \langle\tilde{h}^{\rm obs}_{a,i} \theta_i\rangle_n\)^*
\sum_{j=1}^{\nev} \(\tilde{h}^{\rm true}_{b,j}- \langle \tilde{h}^{\rm obs}_{b,j} \theta_j\rangle_n \) \nn\\
&&\hspace*{-2mm}+\sum_{i,j=1}^{\nev} \[ \langle \tilde{h}^{{\rm obs},*}_{a,i}\theta_i\tilde{h}^{\rm obs}_{b,j} \theta_j\rangle_n 
- \langle \tilde{h}^{{\rm obs},*}_{a,i}\theta_i\rangle_n \, \langle\tilde{h}^{\rm obs}_{b,j} \theta_j\rangle_n \]\nn\\
&&\hspace*{-2mm}-\sum_{i=1}^{\nev} \,\, \langle \theta_i \,(\tilde{n}^*_a\tilde{h}^{\rm obs}_{b,i}+ \tilde{n}_b\tilde{h}^{{\rm obs},*}_{a,i} )\rangle_n 
\, .
\ees
We finally observe that,
for $i\neq j$, in the second line the term $ \langle \tilde{h}^{{\rm obs},*}_{a,i}\theta_i\tilde{h}^{\rm obs}_{b,j} \theta_j\rangle_n $  approximately factorizes into $\langle \tilde{h}^{{\rm obs},*}_{a,i}\theta_i\rangle_n \, \langle\tilde{h}^{\rm obs}_{b,j} \theta_j\rangle_n $, and therefore the contribution of the second line is non-vanishing only when $i=j$.\footnote{The factorization is exact if the events are not overlapping
and  the noise at  the time of arrivals of different events  are uncorrelated. In practice, for BNS, there will be some overlapping signal. However, for each index $i$ in the sum, the sum over $j$ runs over $\nev$ events, where $\nev$ can be of order $10^5$, and of these only a few will have an overlap with the $i$-th signal, so we expect this effect to be small.} Therefore,  our  final form for the noise-averaged correlator is
\bees
&&\hspace*{-8mm}{\neff}_n =\nn\\
&&\hspace*{-4mm}
\sum_{i=1}^{\nev}  \(\tilde{h}^{\rm true}_{a,i}- \langle\tilde{h}^{\rm obs}_{a,i} \theta_i\rangle_n\)^*
\sum_{j=1}^{\nev} \(\tilde{h}^{\rm true}_{b,j}- \langle \tilde{h}^{\rm obs}_{b,j} \theta_j\rangle_n \) \nn\\
&&\hspace*{-4mm}+\sum_{i=1}^{\nev} \[ \langle \theta_i \tilde{h}^{{\rm obs},*}_{a,i}\tilde{h}^{\rm obs}_{b,i} \rangle_n 
- \langle \tilde{h}^{{\rm obs},*}_{a,i}\theta_i\rangle_n \, \langle\tilde{h}^{\rm obs}_{b,i}\theta_i \rangle_n \]\nn\\
&&\hspace*{-4mm}-\sum_{i=1}^{\nev} \,\, \langle \theta_i \,(\tilde{n}^*_a\tilde{h}^{\rm obs}_{b,i}+ \tilde{n}_b\tilde{h}^{{\rm obs},*}_{a,i} )\rangle_n 
\, ,
\label{eq: correlator_subtraction_full} 
\ees
where we recall that all Fourier modes in this equation are evaluated at the same frequency $f$. 

The remaining average over Universe realizations is then automatically taken into account, in the sum over $i=1,\ldots, \nev$, by
extracting the sample of events in accordance with the desired distributions of  parameters. Note that, in principle, also the number of event in the given observation time, $\nev$, is a stochastic variable, and we should  average also over it, with a Poisson distribution. For simplicity, here we keep $\nev$ fixed to the corresponding average value (see also the discussion in Sec.~IV~E of Ref.~\cite{paper:spectraldensity}).

\subsection{Upper bound on the astrophysical confusion noise}\label{sect:Upper}

As discussed above,  the  effect of the astrophysical confusion noise (due to the unresolved sources and to the  error in the subtraction of resolved sources) in a two-detector correlation is not an intrinsic property of the astrophysical background and of the detector network, but also 
depends (through the choice of filter function) on the cosmological background that one is searching. 

In general, it might not be practical to re-evaluate the effect of the astrophysical confusion noise on stochastic searches for each form of the cosmological background that is searched, and it is useful to have statements of more general validity. In particular,
an upper bound on  $(S_{\cal N}/N)_{ab}$   can be obtained if, in \eq{SoverNdelnoiseeff}, rather than using the filter function that maximizes the search for a given cosmological background, we use the filter that maximizes the signal-to-noise ratio $(S_{\cal N}/N)_{ab}$ itself, i.e. that maximizes the effect of the astrophysical confusion noise. Of course, in the search for a cosmological background, the astrophysical confusion noise is a nuisance, and we want it to be as small as possible, not as large as possible. However, computing its effect for the filter function that maximizes it provides an upper bound. In particular, if we  should find that, even with this filter, $(S_{\cal N}/N)_{ab}<1$, then we can forget about the astrophysical confusion noise in all cosmological searches. It should  also be stressed that such an upper bound might turn out to be quite generous: the filter functions used when searching for cosmological backgrounds are smooth functions of frequency: if the spectral density of a cosmological background is a power-law, i.e. $S_h(f)\propto f^{\beta}$, then  $\tilde{Q}_{ab}(f)\propto f^{\beta}/S_{n,ab}^2(f)$; in contrast,  the filter function that optimizes $(S_{\cal N}/N)_{ab}$ will be tuned to the signals from  individual CBCs, so it will be completely different, as we will show explicitly below.

The maximization of $(S_{\cal N}/N)_{ab}$ is performed exactly  as  in \cref{sect:optimal}, with just the appropriate change of notation: the filter that maximizes \eq{SoverNdelnoiseeff} is
\be\label{filtermax}
\tilde{Q}_{ab}(f) \propto \dfrac{\nab^*}{S_{n,ab}^2(f)}\, ,
\ee
and the corresponding maximum value of $(S_{\cal N}/N)_{ab}$  is
\be\label{eq:snrpair}
\snrpair = 
\left[
 \frac{8}{T} 
  \int_{0}^{+\infty} d f \, \dfrac{| \nab |^2}{\Snab^2(f)} 
\right]^{1/2}  \, .
\ee
It is convenient to define a spectral density $\Seffab(f)$ of the astrophysical confusion noise (due to the combined effect of unresolved sources and  the error on resolved sources),  
as
\bees\label{defSconf}
\Seffab(f)&\equiv&\frac{|{\cal N}^{\rm eff}_{ab}(f)|}{|\Gamma_{ab}(f)|}\nn\\
&=&\frac{2}{T}\, \frac{ |\nab |}{|\Gamma_{ab}(f)|}\, .
\ees
From \eq{eq: correlator_subtraction_full}, we see that the terms in the second and third line are given by a single sum over the events  and therefore we expect that they are proportional to the observation time $T$; we also expect that the double sum in the first line will be dominated by the diagonal terms (even more after performing the averages over Universe realizations); in this approximation $\nab\propto T$, so
${\cal N}^{\rm eff}_{ab}(f)$ and $\Seffab(f)$  are (at least approximately) independent of the observation time $T$.

In terms of $\Seffab$, \eq{eq:snrpair} becomes
\be\label{eq:snrSconf}
\snrpair = 
 \[ 2T\int_0^{\infty}df\,  \(\frac{\Gamma_{ab}(f)\Seffab(f)}{S_{n,ab}(f)}\)^2 \]^{1/2}\, ,
\ee
which is analogous to \eq{Vol1_239}.
Similarly to \eq{eq:OmegaSh}, we can define 
\be\label{defOconf}
\oeff(f) \equiv \frac{4\pi^2}{3H_0^2}\, f^3\Seffab(f)\, .
\ee
\Eq{defOconf} provides a first-principle definition of an effective energy density per logarithmic interval of frequency (normalized to $\rho_c$),   due to the combined effect of unresolved sources and the error on resolved sources, to be compared with the energy density carried by a cosmological background.\footnote{Similarly to the discussion  below \eq{defOmeganoise}, it should be stressed that the contribution to  $\oeff(f)$ coming from the error in the subtraction does not correspond, of course, to any physical energy. It is just a contribution to the correlated noise of a detector pair, and therefore to $\Seffab(f)$ that, thanks to \eq{defOconf}, we have expressed in terms of a quantity that can be compared to the actual  energy density  per logarithmic interval of frequency (normalized to $\rho_c$) of a cosmological background.}
We will discuss it further
in \cref{sect:comparison}, where  we will compare it with  related quantities that have been introduced in the literature on more heuristic grounds.

By analogy with \eqs{defhc}{defhn}, we also define (for the case $a\neq b$ that we are considering) the {\em characteristic correlation strain of the astrophysical confusion noise} (after subtraction of the resolved astrophysical sources) as 
\be\label{defhcab}
h_{c,ab}(f)=\[ 2f \Seffab(f)\]^{1/2}\, ,
\ee
and the {\em characteristic  correlation strain of the instrumental noise},
\be\label{defhnab}
\hnab(f)=\[ \frac{2f S_{n,ab}(f)}{|\Gamma_{ab}(f)|} \]^{1/2}\, \frac{1}{(2fT)^{1/4}}\, .
\ee
In terms of them, \eq{eq:snrSconf} reads
\be\label{dlogfhoverh}
\snrpair = \[ \int_{f=0}^{f=\infty} d\log f\, \(\frac{ h_{c,ab}(f) }{ \hnab(f)} \)^4 \]^{1/2}\, .
\ee
Similarly to the definitions in \eqs{defhc}{defhn}, $h_{c,ab}(f)$ and $\hnab(f)$ are dimensionless quantities;  $h_{c,ab}(f)$ carries the information on the ``signal'' that we want to maximize with the two-detector correlation, which in this case is the correlated astrophysical confusion noise, while  we  have put into the definition of $\hnab(f)$  
all quantities  that characterize the detection process (the spectral densities of the uncorrelated noise of the two detectors, their overlap reduction function, and the observation time). Together, $h_{c,ab}(f)$ and $\hnab(f)$ provide spectral information, i.e. tell us from what frequencies $(S_{\cal N}/N)_{ab}^{\rm max}$ gets most of its contribution.

In \cref{sect:rescorr} we will evaluate numerically $h_{c,ab}(f)$ and $\hnab(f)$, as well as $(S_{\cal N}/N)_{ab}^{\rm max}$, for a given 3G network. First, we discuss how to define quantity that carry similar spectral information when the filter is not the one that maximizes $(S_{\cal N}/N)_{ab}[Q]$, which is the case of actual interest when studying the effect of the astrophysical confusion noise in the search for a specific cosmological background.

\subsection{Characterization of the effect of the astrophysical confusion noise}\label{sect:defRab}

In the previous subsection we have computed the effect of the astrophysical confusion noise using the filter function that maximizes it. However,
in the search for a specific cosmological stochastic background, 
the  filter function will be chosen so to maximize the signal-to-noise ratio of that cosmological signal, rather than of the astrophysical confusion noise. The filter function will therefore be given by \eq{Qabopt}, where $S_h(f)$ is the spectral density of the given cosmological signal (that, for simplicity, we assume isotropic and unpolarized). 
Inserting this into \eq{SoverNdelnoiseeff} we get
\be\label{SoverNdSh}
\(\frac{S_{\cal N}}{N}\)_{ab}=\sqrt{2T} \, 
\frac{\left|
\int_{0}^{\infty}df\, {\rm Re} \big[ {\cal N}^{\rm eff}_{ab}(f)  \big] 
\frac{\Gamma_{ab}(f)S_h(f)}{ S_{n,ab}^2(f)}
\right|}
{
\[ \int_{0}^{\infty}df\,  \( \frac{\Gamma_{ab}(f)S_h(f)}{ S_{n,ab}(f)} \)^2 \]^{1/2}
}
\, .
\ee
We observe that, when the filter function was chosen so to maximize the signal-to-noise ratio, the denominator turned out to be just the square root of the numerator. Therefore, we obtain \eq{Vol1_239}, in which $(S/N)^2$ is given simply by an integral over frequencies, and this allowed us to rewrite $(S_{\cal N}/N)_{ab}^{\rm  max}$ as in \eq{dlogfhoverh}, in terms of functions $h_{c,ab}(f)$ and  $\hnab(f)$, that provide information on the relative contribution of different frequency intervals.
This does not happen in \eq{SoverNdSh}, which is in the form of a ratio of two distinct integrals over frequency. Nevertheless, to understand how different ranges of frequencies contribute to the integral, we can rewrite \eq{SoverNdSh}
as
\be\label{SoverNdShdlogf}
\(\frac{S_{\cal N}}{N}\)_{ab}=\left| \int_{f=0}^{f=\infty} d\log f\, R_{ab}(f)\right| \, ,
\ee
where
\be\label{defRab}
R_{ab}(f)=\frac{\sqrt{2T}\, f \, {\rm Re} \big[ {\cal N}^{\rm eff}_{ab}(f)  \big] \frac{\Gamma_{ab}(f)S_h(f)}{S_{n,ab}^2(f)}}{
\[ \int_{0}^{\infty}df'\,  \( \frac{\Gamma_{ab}(f')S_h(f')}{ S_{n,ab}(f')} \)^2 \]^{1/2} 
}
\, .
\ee
Observe that, in \eq{SoverNdShdlogf}, the integrand 
$R_{ab}(f)$ is not positive definite, because at high frequencies $\Gamma_{ab}(f)$ is an oscillating function (although in this regime it is small in absolute value, so, being small and oscillating,  it contributes little to integral over frequency), and also ${\rm Re} [ {\cal N}^{\rm eff}_{ab}(f)]$ is not a priori definite positive. 

The quantity $R_{ab}(f)$ provides spectral information on the frequency range that contribute to $S_{\cal N}/N$. In particular, if, for a given choice of the cosmological spectrum $S_h(f)$,  $|R_{ab}(f)|\ll 1$ over a given frequency range, we can conclude that, in that frequency range, the search for this specific cosmological background is not affected by the astrophysical confusion noise. In \cref{sect:rescorr} we will compute numerically $R_{ab}(f)$ at a 3G detector network, for different choices of $S_h(f)$. 

\section{Source parameters reconstruction beyond Fisher matrix} \label{sect:sourcepar_reconstruction}
We now discuss the procedure that we use for simulating the reconstruction at a detector network of the CBC signals that we inject.
We consider a network of GW detectors, with the detectors labeled by an index $a$, in presence of a GW signal. We  denote by $h_a(t)$ the projection of the GW signal onto the $a$-th detector, obtained as in \eq{hFhF},  and by $n_a(t)$ a realization of the noise in the $a$-th detector. Let $h_a(t,\parset)$ be a waveform template (again, projected onto the $a$-th detector), where $\parset$  collectively denotes the parameters of the template (masses, distance, etc.). Assuming that the GW signal is described by the same template used in the parameter reconstruction, the output of the $a$-th detector can be written as
\be
\label{eq:det_output_signal_true}
s_a(t)=n_a(t) + h_a(t,\trpar)\, ,
\ee
where $\trpar$ denotes the true values of the parameters.
Assuming also that the noise is stationary and Gaussian, and treating the detectors as independent, the log-likelihood for the reconstruction of the signal  parameters is 
\begin{equation}
\label{eq: likelihood}
    -2\log\likelihood{s}{\parset} \propto \sum_a \scalarp{s_a - h_a(\parset)}{s_a - h_a(\parset)} \, ,
\end{equation}
where $s$ on the left-hand side denotes collectively the ensemble of all outputs $s_a$ and, for notational simplicity,  we did not write explicitly the time dependence in $s(t)$, $s_a(t)$ and $h_a(t,\parset)$; in the equation above
 we defined the scalar product between two functions $A(t)$ and $B(t)$  as\footnote{This scalar product, that we denote as $(\,\, |\,\, )_a$, should not be confused with the scalar product $(\,\,\, , \,\,)_{ab}$  defined in \eq{definner}. }
\begin{equation}
\label{eq: scalar product}
    (\, A|B\, )_a = 4 \, \Re\int_{0}^{+\infty} d f \, \dfrac{\ftconj{A}(f) \ft{B}(f)}{S_{n,a}(f)}\,,
\end{equation}
where the tilde denotes the Fourier transform, and $S_{n,a}(f)$ is the PSD of the $a$-th detector, defined by \eq{Sn1}.

Simulating the imperfect subtraction of the astrophysical foreground requires one to distinguish the true signals $\tilde{h}^{\rm true}_a(f)\equiv \tilde{h}_a(f,\trpar)$ from the observed -- or reconstructed -- ones $\tilde{h}^{\rm obs}_a(f)\equiv \tilde{h}_a(f,\obspar)$,
where $\obspar$ are the reconstructed values of the parameters. Such a reconstruction can be performed at different levels of approximation, and different definitions of $\obspar$ are possible.
In many cases in the literature, $\obspar$ is obtained through a Fisher-matrix analysis, to reduce the computational cost of running over large source catalogs needed in order to produce this kind of forecasts.
In this case the likelihood is approximated as a multivariate Gaussian, with covariance given by 
$\CM_{\alpha\beta}(\trpar)=\left(\FM^{-1}\right)_{\alpha\beta}(\trpar)$, where
\begin{equation}
    \FM_{\alpha\beta}(\trpar) = \sum_a \scalarp{\de{\alpha} h_a(\trpar)}{\de{\beta} h_a(\trpar)} \, ,
\end{equation}
and the indices $\alpha, \beta$ label the parameters of the waveform.
Specifically, after computing and inverting a Fisher matrix for each event, the observed parameters $\obspar$ are estimated by extracting a sample from the associated multivariate Gaussian likelihood, often restricted to a subspace of the whole parameter space. However, as already pointed out in Ref.~\cite{Sharma:2020btq}, this approach is not properly rooted in the spirit of a close-to-reality observation since, in this way, the reconstruction is not related to a specific noise realisation of the detector. 
As discussed in Refs.~\cite{Vallisneri:2007ev,Iacovelli:2022bbs}, one can take into account  the impact of the specific noise realization, while still relying on Fisher-matrix analyses, defining $\obspar$ as 
 the  maximum likelihood  estimator, that we denote as $\mlpar$; the latter satisfies  (to linear order in an expansion in $1/{\rm SNR}$) 
\begin{equation}
\label{eq: parameter shift}
    \theta^{\rm ML}_{\alpha} - \theta^{\rm true}_{\alpha} \simeq \CM_{\alpha\beta} (\trpar) \sum_a \scalarp{ n_a }{ \de{\beta} h_a(\trpar)  } \, .
\end{equation}
However,  the use of \eq{eq: parameter shift}  can still result in artificially large subtraction errors, as the Fisher matrix approximation is known to fail in  regions of the parameter space where degeneracies are important.\footnote{\label{footnote:FIM_fail_dLi}As an example, the Fisher approach is known to fail in the luminosity distance-inclination plane for face-on/face-off systems, overestimating the statistical uncertainty on these parameters. This could result in extracting a sample considerably far from the true value, producing an artificially high subtraction error. See, however, Ref.~\cite{Mancarella:2024qle} for techniques for dealing analytically with this specific degeneracy.} On the other hand, if one  restricts the analysis to a parameter space of too small dimension, the contribution from imperfect subtraction risks to be severely underestimated, as shown in Refs.~\cite{Zhou:2022otw,Zhou:2022nmt}.

For these reasons, in this work we adopt a different strategy not relying on the Fisher-matrix approach: 
for a given noise realization, we directly numerically search for the maximum of the log-likelihood (\ref{eq: likelihood})
as a function of the source parameters $\parset$. As we discussed in \cref{sect:Astrocorrelated noise}, the effect of the residual astrophysical foreground on the detectability of cosmological backgrounds is contained in the correlator given in \eq{eq: correlator_subtraction_full2}.
The quantity inside this correlator  depends on the noise realization both explicitly, see the last  line of \eq{eq: correlator_subtraction_full2}, and through the maximum likelihood values of the parameters, and will be computed by averaging over many noise realizations. Although computationally more expensive than Fisher-based methodologies, this procedure leads to more realistic results, does not rely on any approximation of the likelihood, and can easily incorporate physical priors on the parameters. 
Moreover, its cost can be mitigated by initializing the maximizer in a region of the parameter space close to the injected parameters.

We model the frequency-domain emissions with the \texttt{IMRPhenomXHM}~\cite{Pratten:2020fqn,Garcia-Quiros:2020qpx} waveform template for BBHs, which includes higher-order modes, 
while for BNSs we use  \texttt{IMRPhenomD\_NRTidalv2}~\cite{Husa:2015iqa,Khan:2015jqa,Dietrich:2019kaq}, which includes tidal effects.\footnote{We stress that the use of time-domain approximants in \cref{fig:src_strain_sim} is only for illustrative purposes, and these models are not used in the analysis, which is carried out in the frequency domain.} 
The inference is therefore performed on a large parameter space (the same used in \cite{Iacovelli:2022bbs,Branchesi:2023mws}): the parameters of these waveforms are 
\be
\{ {\cal M}_c, \eta, d_L, \theta, \phi, \iota, \psi, t_c, \Phi_c, \chi_{1,z}, \chi_{2,z}, \tilde{\Lambda}\}\, ,
\ee
where ${\cal M}_c$ is the detector--frame chirp mass, $\eta$ the symmetric mass ratio, $d_L$ the luminosity distance to the source, $\theta$ and $\phi$  the sky position coordinates, $\iota$ the  angle between the orbital angular momentum of the binary and the line of sight, $\psi$ the polarisation angle, $t_c$ the time of coalescence, $\Phi_c$ the phase at coalescence, $\chi_{i,z}$ the dimensionless spin of the object $i=\{1,2\}$ aligned with the orbital angular momentum, and $\tilde{\Lambda}$ the 5-PN combination of the dimensionless tidal deformability of the objects~\cite{Wade:2014vqa} (which is present only for systems containing a NS).

Using a large parameter space is important because, as shown in Refs.~\cite{Zhou:2022otw,Zhou:2022nmt}, using a restricted subsample of parameters results in  over-estimating the actual reconstruction capabilities of a detector network, eventually leading to incorrect estimates of the cumulative error  on resolved sources, even at the level of orders of magnitude.

We project these waveforms over the  whole frequency band probed by the considered 3G detectors, through the publicly available software \texttt{GWFAST}~\cite{Iacovelli:2022mbg}.
We use the same waveform for the injections and for the recovery of the signals. Waveform systematics will in principle further contribute to the reconstruction error, although  this will also depend on  future progresses in constructing more and more faithful templates at the level of accuracy needed by 3G detectors; see 
Refs.~\cite{Kapil:2024zdn,Dhani:2024jja} for recent discussions.

\section{Numerical results}\label{sect:rescorr}

We now evaluate numerically the effect  of the astrophysical confusion noise and its spectral properties, in a selected 3G network. As we have repeatedly stressed, the effect of the astrophysical confusion noise also depends on the cosmological search that is performed, through the choice of the optimal filter function, and we will examine several different cases. We will first compute the results  for the  filter function that maximizes $(S_{\cal N}/N)_{ab}$ itself, i.e. that maximizes  the astrophysical confusion noise,
providing in this way an absolute upper bound on its effect. We will then turn to the effect of the astrophysical confusion noise on the search for some selected  cosmological backgrounds,  by evaluating numerically  the quantity $R_{ab}$ defined in \cref{sect:defRab}. We will first evaluate it when the filter optimizes the search for some selected power-law  cosmological backgrounds,  and we will then compute it when the filter 
is chosen so to optimize the search for cosmological spectra corresponding to some broad bump, as in the case of cosmological phase transitions.

\subsection{3G detector network}\label{sect:3Gnetwork}

In this work we consider just one example of a 3G detector network, made by ET in its 10~km triangle configuration, located in Europe, plus 2 CE detectors, one of 40~km and one of 20~km arm-length, both located in the US. In particular, we use the same detector locations and orientations, as well as the same PSDs, as in Ref.~\cite{Branchesi:2023mws}. Currently, different options for 3G detectors are under study: in particular, for ET an alternative to the triangle configuration, under active consideration,  is given by two well-separated L-shaped detectors, both located in Europe~\cite{Branchesi:2023mws}. Given that our analysis is computationally quite demanding,  we limit here to this single 3G detector network, with the main aim of presenting the methodology. We defer to further work a comparison of the effect of the residual astrophysical background on different network configurations.

In its triangular configuration, ET is made of three nested detectors (with each detector composed of two interferometers, one tuned toward low-frequencies and one toward high-frequencies), that we will denote here as ${\rm ET}_1$, ${\rm ET}_2$ and ${\rm ET}_3$. However, these three outputs are not independent and one combination (the sum of the three outputs) is the so-called null-stream, where the GW signal cancels (see Ref.~\cite{Goncharov:2022dgl} and Sec.~7 of Ref.~\cite{Branchesi:2023mws} for a discussion of the potential and the limitations of the null stream for data analysis at ET). For the purpose of counting independent detectors, ET in the triangular configuration therefore has two independent detectors, that we will take to be 
${\rm ET}_1$ and ${\rm ET}_2$. Actually, for the triangular configuration of ET, different linear combinations, involving all three detectors, can be constructed to search for GWs and to form the null stream. For instance for LISA~\cite{LISA:2017pwj}, denoting the three TDI channels by $X,\, Y,\, Z$, one considers the combinations $A = (2 X - Y - Z)/3$ and $E = (Z-Y)/\sqrt{3}$, while the null channel is $T = (X + Y + Z)/3$, see e.g. Refs.~\cite{Prince:2002hp, Adams:2010vc}. These have the advantage of being noise-orthogonal in the symmetric noise limit. For our purposes, it will be sufficient to show the results in the simpler ${\rm ET}_1$ and ${\rm ET}_2$ basis.

Together with CE-40km and CE-20km, the network that we are considering therefore has, overall, four independent detectors. The full detector network will be used to 
detect and reconstruct the injected CBC signals, 
and we will then evaluate the effect of the astrophysical confusion in a two-detector correlation,   for each pair of detectors.
With four independent detectors (${\rm ET}_1$, ${\rm ET}_2$, CE-40km and CE-20km) we can form 6 detector pairs.
We will then show the results for the four pairs:
\begin{itemize}[label=--]
\item (${\rm ET}_1$, ${\rm ET}_2$)\,;
\item (${\rm ET}_1$, CE-20km)\,;
\item (${\rm ET}_1$, CE-40km)\,;
\item (CE-40km, CE-20km)\,.
\end{itemize}
The results for (${\rm ET}_2$, CE-20km) and (${\rm ET}_2$, CE-40km) are essentially the same as for
(${\rm ET}_1$, CE-20km) and (${\rm ET}_1$, CE-40km), respectively.

\subsection{Technical aspects}

Independently of the filter function chosen, the first step  in evaluating $(S/N)_{ab}[Q]$  from \eq{SoverNdelnoiseeff} (and the one which is  computationally very intensive)  is the computation of 
the correlator $\left\langle \tilde{n}^{{\rm eff},*}_a(f)\tilde{n}^{\rm eff}_b(f) \right\rangle_n$, averaged over many noise realizations. Given the injected values 
$\tilde{h}^{\rm true}_{a,i}$ and a noise realization, we compute the observed value
$\snrobs$ of the signal-to-noise ratio of each CBC of our catalog in the given noise realization;  if $\snrobs$ is larger than a chosen threshold $\snrth$, we reconstruct the signal according to the procedure discussed in \cref{sect:sourcepar_reconstruction},  obtaining the corresponding value of 
$\tilde{h}^{\rm obs}_{a,i}$. We perform this operation for all the  CBCs in our catalog, $10^4$ BBHs and $10^4$ BNSs, obtained extracting a random subset from the BBH and BNS catalogs used in Ref.~\cite{Branchesi:2023mws}, which was generated 
using the most recent population synthesis models, combined with the LVK observations.\footnote{For the BNS case, we use the catalog featuring a Gaussian mass distribution of the sources, and adimensional tidal deformabilities computed assuming the APR equation of state~\cite{Akmal:1998cf}.} We then repeat the procedure for $\nnoise = 10^3$ different noise realizations which,   performing tests with different numbers of noise realizations, turned out to be  a good trade-off between computational cost and convergence of the results. This means that  each of the $10^4$ BBH  and each of the $10^4$ BNS of our catalogs is reconstructed for $10^3$ different realizations of the noise, using the reconstruction procedure detailed in \cref{sect:sourcepar_reconstruction}, which involves a full numerical maximization of the likelihood in the given noise realization, rather than just a simple Fisher-matrix approximation, leading  to a rather computationally expensive analysis.
We finally repeat the whole process for several  different values of $\snrth$.

The values
$N_{\rm BBH}=N_{\rm BNS}=10^4$ come from a compromise between having a statistically significant sample, and keeping manageable the computational cost (given that each source will be combined with $10^3$ different noise realizations). 
The population model used in Ref.~\cite{Branchesi:2023mws}, that we are also using here, corresponds to about $1\times 10^5$ BBH 
and $7\times 10^5$ BNS coalescences in one year, so in this population model our sample corresponds to 36 days of observations  for BBHs and 5 days for BNS.\footnote{More precisely, this is the  number of days of  simultaneous data taking of the two detectors, in each two-detector pair. The  actual data taking time will also depend on the duty cycle of the detectors. E.g., assuming an independent $85\%$  duty cycle in each detector, 36 days of coincident observations would correspond to about 50 days of actual time in a two-detector pair.}
However, these numbers are of course subject to uncertainties. The current uncertainty on the local merger rate of BBHs is about a factor of 3, and the evolution of their merger rate density with redshift at $z\gsim1$, is still unconstrained by the current data~\cite{KAGRA:2021duu}. For BNSs the situation is even more uncertain, as observations of these kind of systems are more rare. In particular, the current uncertainty on their local merger rate given by LVK data spans three orders of magnitude, and again the merger rate density is poorly known. Using O3 data, the number of BNS coalescences in one year could thus be as low as ${\cal O}({\rm few}\times10^4)$ or as high as a ${\cal O}({\rm few}\times10^6)$~\cite{KAGRA:2021duu}. 
The fact that, to date, no BNS candidate has been announced in the ongoing O4 run brings the BNS rate closer to the lower end of this range.
We will discuss below how to approximately rescale our results for different number of sources.

To generate the noise realizations, 
given a detector network, 
we introduce a (linear) grid in frequency space, with spacing $\Delta f = \SI[parse-numbers=false]{2^{-3}}{\hertz}$,  which ranges from \SI{2}{\hertz} to \SI[parse-numbers=false]{2048}{\hertz}.
Assuming the noise to be stationary and Gaussian, for each detector of the network we extract the individual complex realizations $\tilde{n}_a(f) = \Re \,\tilde{n}_a(f) + i \, \Im \, \tilde{n}_a(f)$ from the associated probability distributions
\begin{equation}
        p\left( \Re \, \tilde{n}_a(f) \right) = p\left( \Im \, \tilde{n}_a(f) \right) = \normdist{\dfrac{S_{n,a}(f)}{4 \Delta f}} \, ,
\end{equation}
where $\mathcal{N}\left( \mu, \sigma^2 \right)$ denotes a normal distribution with mean $\mu$ and variance $\sigma^2$.\footnote{We  will assume  that the  noise of the different detectors are uncorrelated, even in the case of a triangular-shape interferometer. The effect of correlated noise on parameter estimation of CBCs at 3G detectors has been recently studied in Ref.~\cite{Wong:2024hes}, that presents the results for a whole range of values of the correlation coefficient, including very high values. When using  coherence values such as those found in Ref.~\cite{Janssens:2024jln} from actual measurements of the correlation of the underground seismic field, the results of Ref.~\cite{Wong:2024hes} indicate that, for parameter estimation of coalescing binaries, the effect of correlated noise is in fact not very  relevant.
Note also that the results in Ref.~\cite{Wong:2024hes} only apply to a hypothetical configuration of two colocated L-shaped detectors, and not to the triangle configuration.
}

\subsection{Results for the upper bound on the astrophysical confusion noise}\label{sect:resultsupper}

\begin{figure*}
    \centering
    \includegraphics[width=0.8\linewidth]{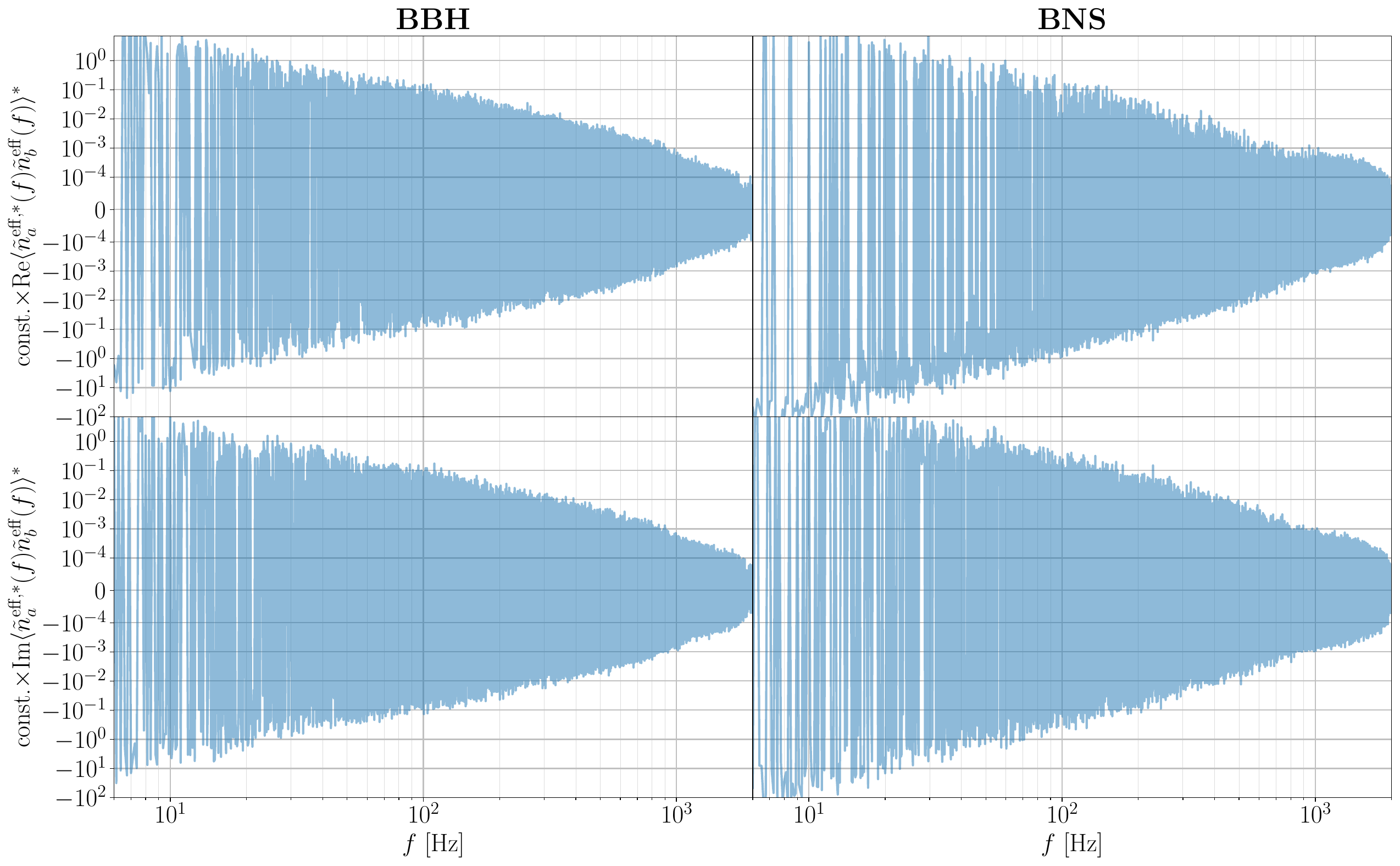}
    \caption{The real and imaginary parts of 
    $\langle\tilde{n}^{\rm eff, *}_a(f)\tilde{n}^{\rm eff}_b(f)\rangle^*$, for the ET$_1$-ET$_2$  pair, setting $\snrth = 12$.}
    \label{fig:max_filter}
\end{figure*}

We begin with the case, discussed in \cref{sect:Upper}, where the filter function is chosen so to maximize the effect of the astrophysical confusion noise on a two-detector correlation, and which therefore only provides an upper bound on its actual relevance. 

The filter that maximized the astrophysical noise is given by \eq{filtermax} [with the correlator given by \eq{eq: correlator_subtraction_full}], while 
the one that maximizes a cosmological background with spectral density $S_h(f)$ is given by \eq{Qabopt}. For cosmological backgrounds, within the bandwidth of ground-based detectors,  $S_h(f)$ is in general a smooth function. This is due to the fact that typical cosmological spectra extend  across many  decades in frequencies, so the bandwidth explored by ground-based detectors is, comparatively, small. In many cases, within this bandwidth, the predicted cosmological spectra can be well approximated by a power-law behavior $S_h(f)\propto f^{\alpha}$, or else (as in the case of phase transitions) by a function with a relatively broad peak. At the same time, in the region of frequencies that contributes significantly to the signal-to-noise ratio, the overlap reduction function in \eq{Qabopt} is a smooth function of order one.  To have an intuitive understanding of how much  the filter function (\ref{filtermax}) can differ from the cosmological filters (\ref{Qabopt}), 
it is interesting to plot the real and imaginary parts of $\nab^*$, to see how it compares with a smooth function such as $\Gamma_{ab}(f)S_h(f)$. In both cases, the filter is then eventually obtained dividing by $S_{n,ab}^2$, which weights more the part of the spectrum where the detector sensitivities are better.

Fig.~\ref{fig:max_filter} shows the real and imaginary parts of $\nab^*$, obtained setting $\snrth=12$, for the ET$_1$-ET$_2$ pair. We see that, both for BBHs and BNSs, the real and imaginary parts have qualitatively similar behaviors, characterized by strong oscillations. The filters that optimize the contribution of the astrophysical confusion noise are therefore very different from those that optimize the search for cosmological background and we can thus expect that, when we compute the correlation of the astrophysical noise with the filter optimized to cosmological searches, their effect will be significantly smaller that the upper bound that we will compute in this section. 

\begin{figure*}[tbp]
    \centering
    \includegraphics[width=.94\linewidth]{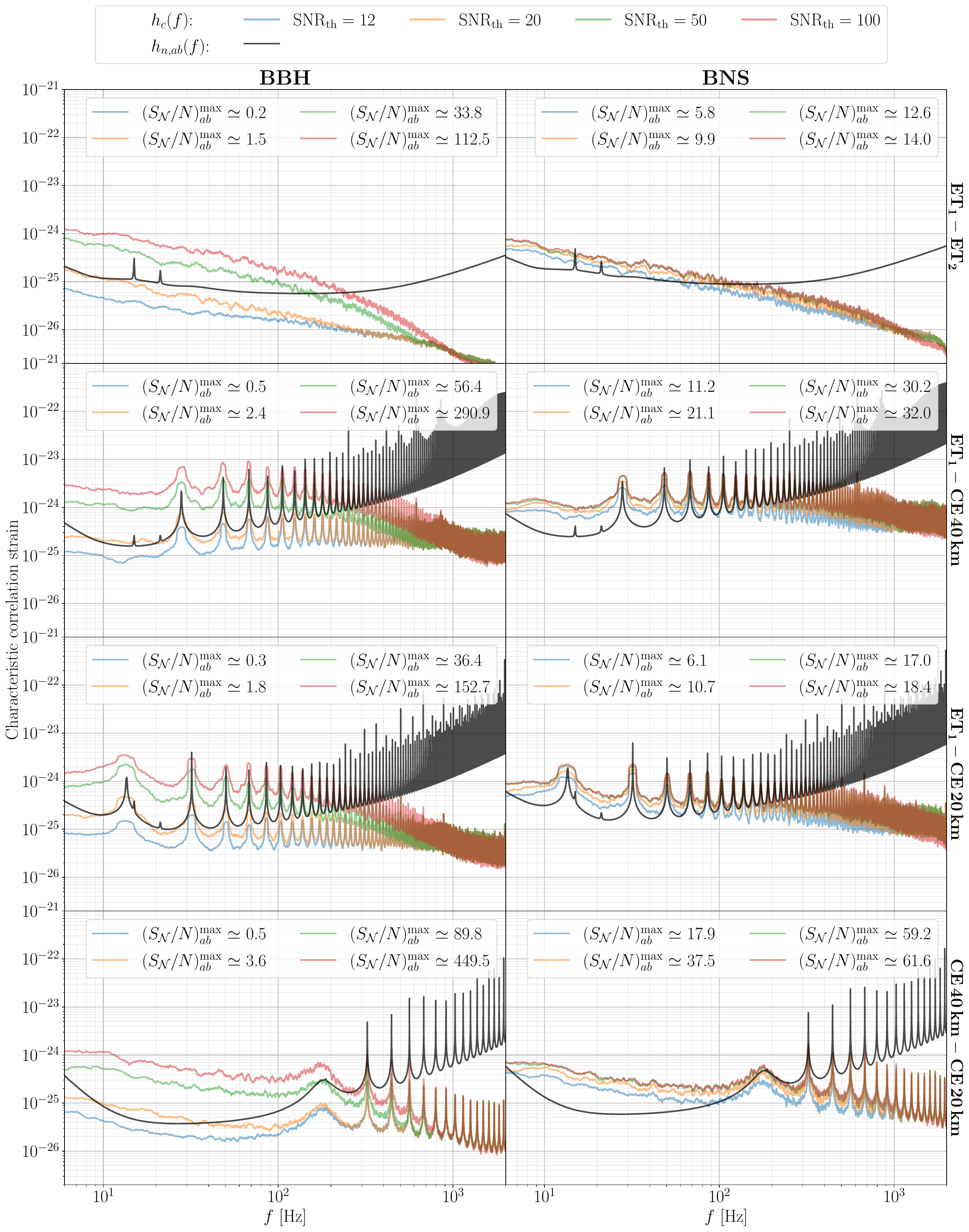}
    \caption{The
    characteristic correlation strain of the astrophysical confusion noise, $\hcab(f)$, defined in \eq{defhcab}, computed assuming different values of the threshold $\snrth$ that defines when a CBC is deemed detected and  is subtracted (colored lines, colours as in the  legend), compared to
    the characteristic correlation strain of the instrumental noise, $\hnab(f)$, defined in \eq{defhnab} (black lines),  
    for a catalog of $10^4$ BBHs (left panels) and $10^4$ BNSs (right panels). Each row corresponds to one of the  four detector pairs considered. The corresponding values of $(S_{\cal N}/N)^{\rm max}_{ab}$ , computed as in \eq{dlogfhoverh}, are shown in the insets. }
    \label{fig:subtraction_test_different_snrth}
\end{figure*}

The results for the upper bound are shown in \cref{fig:subtraction_test_different_snrth}. 
On the left panels we show the result for $10^4$ BBHs, while on the right panels we show those for $10^4$  BNSs. The four rows correspond to the four detector pairs listed in \cref{sect:3Gnetwork}. The black line is  
the characteristic correlation strain of the instrumental noise, $\hnab(f)$, defined in \eq{defhnab}. Observe that it diverges at the zeros of the overlap reduction function; these depend strongly on the distance between the two detectors, so the corresponding series of peaks in $\hnab(f)$ starts already at around 30~Hz  for the ET-CE pairs, that have an intercontinental baseline, around 300~Hz for the CE-CE pair, where we have located both CE detectors in the US (of course, the detailed position of the peaks depends on the  location chosen within the US), and is absent for the colocated ET-ET pairs.\footnote{Actually, given the distance ${\cal O}(10)$~km between the vertices of the different detectors making ET in the triangle configuration, the  overlap reduction function between two ET detectors  decrease and undergoes oscillations starting from frequencies of order
$f\sim c/10~{\rm  km}\sim 30~{\rm kHz}$. In contrast, the two small peaks visible at low frequencies are due to the  shape of $S_{n,\rm ET}(f)$.}
The colored lines are the
characteristic correlation strain of the astrophysical confusion noise, $\hcab(f)$, defined in \eq{defhcab}, computed for  different values of $\snrth$, the threshold that determines whether a CBC is considered resolved (and is then subtracted) or is left unresolved.
Recall that the  integrand of \eq{dlogfhoverh}
is $[\hcab(f)/\hnab(f)]^4$. Therefore, at frequencies for which $\hcab(f)\ll \hnab(f)$, the contribution to the two-detector correlation of the astrophysical confusion noise stays below the contribution of the instrumental noise, even in this limiting case  in which we are using the filter function that maximizes the signal-to-noise ratio of the astrophysical confusion noise.
The corresponding value of the integral, i.e. of $(S_{\cal N}/N)_{ab}^{\rm  max}$, is also shown in the inset of each plot.

To understand the meaning of these results let us recall, first of all, that they have been obtained for a fixed number of sources (detected or undetected). We have defined $\hcab(f)$ and $\hnab(f)$  so that $\hcab(f)$ is independent of the observation time, while $\hnab(f)$
decreases with time as $1/T^{1/4}$, see \eqs{defhcab}{defhnab}. Therefore, 
\be
\hnab(f)\propto \frac{1}{N^{1/4}}\, ,
\ee
where $N$ is the number of events of the given type (BBHs or BNSs, respectively) coalescing in the observation time $T$.
Therefore, to understand whether, over a given observation time, the correlated astrophysical confusion noise, expressed by $\hcab(f)$, can become larger than the instrumental noise $\hnab(f)$, one must  rescale the  results for $\hnab(f)$ in \cref{fig:subtraction_test_different_snrth} 
by a factor $(10^4/N_{\rm BBH})^{1/4}$ for BBHs, and $(10^4/N_{\rm BNS})^{1/4}$ for BNSs, where $N_{\rm BBH}$ and $N_{\rm BNS}$ are the number of BBH and, respectively, BNS coalescences taking place in the Universe during the given observation period.\footnote{Of course, this is just an approximated procedure, valid if the sample of $10^4$ coalescences that we have used is, statistically, sufficiently representative; increasing the size of a sample always brings out more rare events, which are in the tails of the distributions, and that cannot be simply accounted for by a rescaling. The rigorous approach would be to repeat our analysis with the desired number of events.}
Because of the $1/4$ power, however, the results for $\hnab$ do not have a very strong dependence on the number of events,  and will in general not differ too much from those that we  show with our reference values $N_{\rm BBH}=N_{\rm BNS}=10^4$. For instance,  setting $N_{\rm BNS}=10^5$, the result for $\hnab(f)$ would only change by a factor $10^{1/4}\simeq 1.8$, while for 
$N_{\rm BNS}=10^6$, it would  change by a factor $\sqrt{10}\simeq 3.2$. Recall, however, that  the corresponding values of $(S_{\cal N}/N)^{\rm max}_{ab}$  increase as   $N_{\rm BBH}^{1/2}$ and $N_{\rm BNS}^{1/2}$, respectively.

With this understanding, we can now discuss the results shown in \cref{fig:subtraction_test_different_snrth}.
Our first comment is on the dependence of the results on $\snrth$, the threshold above which a CBC is defined as detected and its contribution is subtracted, while below it is left unresolved. We see that, the lower this threshold, the better is the result, in the sense that the astrophysical confusion noise decreases monotonically as we lower $\snrth$, and we get a correspondingly lower value of the signal-to-noise ratio  of the astrophysical confusion noise. The result is  evident for BBHs, while for BNSs the effect is smaller and there is a tendency toward saturation of the different curves, but still the behaviour is monotonic.  This  means that, even when a CBC  is detected at relatively low  signal-to-noise ratio, and therefore in general its parameters are not very well reconstructed, still  subtracting it from the detectors' outputs is better than leaving it as unresolved.
This makes sense:
having some information and using it to clean  the output is  better than having no information at all. 
Unless the subtraction of low-SNR sources introduces a systematic bias, 
we expect that, at most, the improvement saturates when we begin to remove sources that are too poorly measured, so that subtracting them, or leaving them unresolved, becomes equivalent.\footnote{For BBHs, we studied the dependence on the $\snrth$ down to $\snrth=8$, and we find that the curves for $\snrth=12$ and $\snrth=8$ are basically superimposed, within statistical fluctuations,  in the whole frequency range, so decreasing $\snrth$ from 12 to 8 there is no further gain, and the results saturate. Note that the monotonic  dependence on  $\snrth$  that we find is different from that found in Refs.~\cite{Zhou:2022otw,Zhou:2022nmt}. In \cref{sect:comparison} we will compare our approaches.}

From the left panels of \cref{fig:subtraction_test_different_snrth} we see that, for the specific 3G network that we are considering, and an observation time corresponding to $N_{\rm BBH}=10^4$, even in the extreme case that we are considering, where we have chosen the filter that maximizes the signal-to-noise ratio of the astrophysical confusion noise,  the contribution of
BBHs to the two-detector correlation can be pushed  below that of instrumental noise. We see that (for our detector network, and $N_{\rm BBH}=10^4$) already removing the BBHs with $\snr \geq20$ is sufficient to bring the upper bound on the BBH confusion noise below the instrumental noise for all detector pairs considered (except for the CE40km-CE20km pair where, for  BBHs, this upper bound   stays above the instrumental noise  in a frequency window between about 10 and 30~Hz), while removing the BBHs with $\snr \geq12$
brings the upper bound on the BBH confusion noise well below the instrumental noise, everywhere in frequency, in all detector pairs; the corresponding values of $\snconf$ are between 0.2 and 0.5 in the various detector pair. Raising $N_{\rm BBH}$ to $10^5$, that in our catalog correspond to 1 year of simultaneous data taking, 
$\snconf$ increase by a factor $\sqrt{10}\simeq 3.2$, and remains around or below 1 in each detector pair. Setting a safer detection threshold at $S/N=3$, the upper bound on the BBH confusion noise will remain below the threshold even integrating for ${\cal O}(10)$~yr. 

Recalling that  $h_{n,ab}$ provides an upper bounds on the actual  BBH confusion noise, obtained with the filter function that maximizes it, rather than with the filter functions actually used in cosmological searches, we can  conclude 
that, at this 3G network, the BBH confusion noise can be subtracted so that it does not affect the 
sensitivity of  to cosmological stochastic backgrounds. 

The situation is potentially different for BNSs. As we see from the right panels, subtracting all sources with $\snr \geq12$
brings the BNS confusion noise below the instrumental noise  above  about $10^2$~Hz, but at $f \,\lsim\, 100$~Hz it stays above, in all detector pairs of the network that we are considering, and the corresponding value of $(S/N)_{ab}^{\rm max}$, obtained integrating over  frequencies, \eq{dlogfhoverh}, is in the range 6 to 18, depending on the specific detector pair.
Once again, what we are showing here is an upper bound on the actual BNS confusion noise, obtained with the filter function that maximizes it, rather than with the filter functions actually used in cosmological searches.
Therefore, a first conclusion is that, at this 3G network,  BNSs do not spoil the search for cosmological stochastic backgrounds at $f\,\gsim\, 100$~Hz. They could potentially   have an impact at $f\, \lsim\, 100$~Hz but, to assess  whether this actually happens, we must move from the evaluation of the upper bound to a realistic estimate using typical smooth filter functions appropriate to the searches for cosmological stochastic backgrounds.
We will address this issue in the next subsection.

We emphasize, however, that the above conclusions strongly depend  on the  detector network considered, since this crucially determines the quality of the reconstruction and subtraction of the resolved sources.
Generally speaking, we expect that in a less performant network, such as that made for instance just by a single ET triangular detector, or  just  by two L-shaped ET detectors both in Europe, or else  by just two CE detectors in the US, the effect of the astrophysical confusion noise will be much more important. We defer a more systematic study of the effect on different 3G networks to further work.

\subsection{Results for power-law cosmological backgrounds}\label{sect:resultsPL}

We next compute the effect of the astrophysical confusion noise  in the search for power-law cosmological backgrounds, 
using now the filter functions appropriate to such searches,  rather than the filter that maximizes the effect 
of the astrophysical noise itself.

\begin{figure*}[tbp]
    \centering
    \includegraphics[width=.96\textwidth]{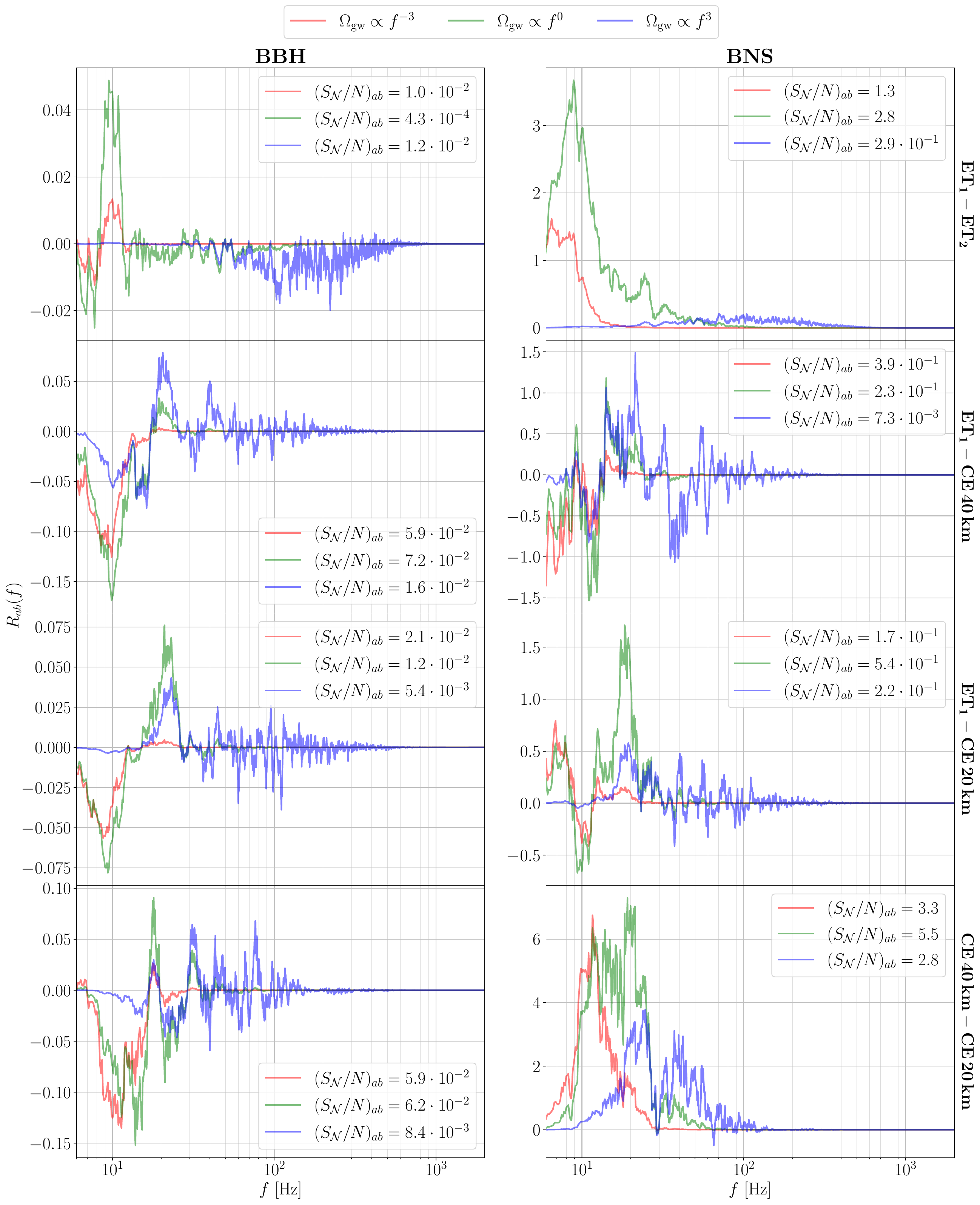}
    \caption{The function $R_{ab}(f)$ that determines the signal-to-noise ratio of the astrophysical confusion noise through \eq{SoverNdShdlogf}, for three different cosmological searches corresponding to $\omgw(f)\propto 1/f^3$,
    $\omgw(f)\propto {\rm const.}$, and $\omgw(f)\propto f^3$. The inset gives the quantity $( S_{\cal N}/N)_{ab}$, defined in \eq{SoverNdShdlogf} as the integral of  $R_{ab}(f)$ over $d\log f$. We use $\snrth=12$ as threshold for the resolved sources. 
    }
    \label{fig:R_ab}
\end{figure*}

In \cref{fig:R_ab} we show the result  for $R_{ab}(f)$, defined in 
\eqs{SoverNdShdlogf}{defRab}, for power-like spectra $\omgw(f)\propto f^{\alpha}$ and $\alpha=-3,0,3$.
For BBHs, the results shown in the left column confirm those already obtained from the PLS and from the upper bound. We see, in particular, that the values of $(S_{\cal N}/N)_{ab}$ (given in the inset for each detector pair) are small, at most of order a few times $10^{-2}$. The right column gives the results for BNS. The results  confirm that, for the (${\rm ET}_1$, CE-20km) and (${\rm ET}_1$, CE-40km) pairs, the BNS confusion noise produces a value of $(S_{\cal N}/N)_{ab}<1$ and therefore does not spoil the instrumental sensitivity. For the other two pairs,  (${\rm ET}_1$, ${\rm ET}_2$) and (CE-40km, CE-20km), \cref{fig:R_ab} shows that 
$(S_{\cal N}/N)_{ab}>1$, although only by factors of a few, 
and also shows that basically no contribution to $(S_{\cal N}/N)_{ab}$ comes from the region $f>100$~Hz. The most significant contribution in general come from the region $f\sim 10-30$~Hz, except  for the (${\rm ET}_1$, ${\rm ET}_2$)  pair, where it is dominated by the region $f<10$~Hz.

\subsection{Results for cosmological backgrounds from phase transitions}
\label{sect:PT}

An important example of spectra of cosmological origin that cannot be described by a single power-law is given by those produced during phase transitions in the early Universe, which typically are rather described by   broad bumps, possibly with superimposed features due to different concurring mechanisms such as bubble collisions, magneto-hydrodynamics turbulence, or sound waves~\cite{Maggiore:2018sht,Caprini:2018mtu}. 
However, a
useful feature  of the quantity $R_{ab}(f)$ introduced in \cref{sect:defRab} is that it can be used for arbitrary spectral shapes. In this section we evaluate it for a cosmological background featuring a broad bump. As an example, we consider GWs produced by bubble collisions during first-order phase transitions. 
An analytic fit to the most recent  numerical simulations gives~\cite{Caprini:2018mtu}
\be\label{omgwbubbles}
\Omega_{\rm gw, bc}(f) = C_{\rm bc}\, \frac{(f/f_{\rm bc})^{\alpha}}{1 +\alpha (f/f_{\rm bc})^{1+\alpha}}\,,
\ee
where $\alpha\simeq 2.8$, and the overall constant $C_{\rm bc}$ and the peak frequency   $f_{\rm bc}$ can be computed in terms of the underlying parameters of the phase transition (the subscript `bc' stands for `bubble collisions'). In particular, $f_{\rm bc}$ depends on the temperature $T_*$ at which the  phase transition takes place, and falls in the ET bandwidth, between a few Hz and a few kHz,  for $T_*$ between ${\cal O}(10^7)\, {\rm GeV}$ and
${\cal O}(10^{10})\, {\rm GeV}$~\cite{Maggiore:1999vm}; a strong first-order phase transition at such scales could for instance occur naturally in axion models with a spontaneously broken Peccei-Quinn symmetry~\cite{Caprini:2024ofd}. Note that
the spectrum in \eq{omgwbubbles} grows as $f^{2.8}$ below the peak frequency, and decreases as $f^{-1}$ above.

\begin{figure*}
    \centering
    \includegraphics[width=1.\linewidth]{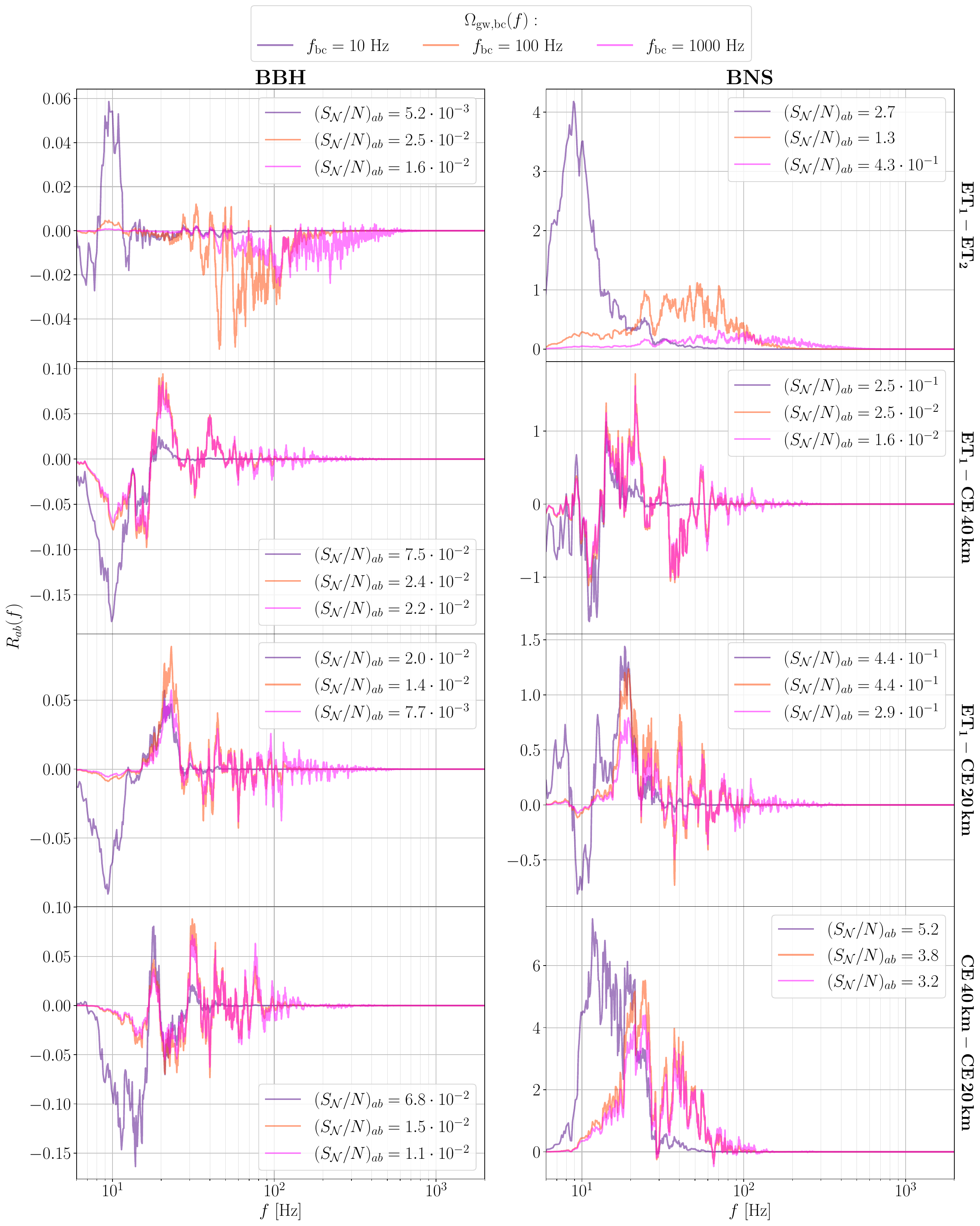}
    \caption{The function $R_{ab}(f)$, describing the astrophysical confusion noise, in the search for a cosmological background generated by bubble collisions during a first order phase transition in the early Universe, parametrized as in \eq{omgwbubbles}, for three different values of the peak frequency, $f_{\rm bc}=10, 100$ and 1000~Hz.
    We use again $\snrth=12$ as threshold for the resolved sources. }
    \label{fig:r_ab_nopowerlaw}
\end{figure*}

Qualitatively similar spectra are produced by other mechanisms associated to phase transitions; e.g., sound waves produce a spectrum fitted by~\cite{Caprini:2018mtu}
\be\label{omgwsw}
\Omega_{\rm gw, sw}(f) = C_{\rm sw}\, \frac{(f/f_{\rm sw})^{3}}{[4 +3 (f/f_{\rm sw})^2\, ]^{7/2}}\,,
\ee
i.e. growing as $f^3$ at frequencies well below the peak at $f=f_{\rm sw}$, and decreasing as $f^{-4}$ at $f\gg f_{\rm sw}$,
while magneto-hydrodynamics turbulence produces a spectrum~\cite{Caprini:2018mtu}  
\be\label{omgwturb}
\Omega_{\rm gw, turb}(f) = C_{\rm turb}\, \frac{(f/f_{\rm turb})^{3}}{[1 +(f/f_{\rm turb})]^{11/3} (1+8\pi f/f_*)}\,,
\ee
(where $f_*$ is the frequency corresponding to the Hubble parameter at production, redshifted to today), so it 
grows again as $f^3$ at frequencies well below the peak and, at sufficiently large $f$, decreases as $f^{-5/3}$.

In a realistic study of the detectability of the signal from a phase transition, all these contributions should be added, resulting in a structure with multiple bumps.
Here we are just interested in a more generic methodological question, i.e. how well one can subtract the astrophysical confusion noise when searching for a cosmological background featuring a broad bump and, for definiteness, we will use the functional form given in \eq{omgwbubbles}.

The corresponding optimal filter function is obtained as usual from \eq{Qabopt}, inserting the corresponding expression for $S_h(f)$. Note that the overall normalization of the spectrum, i.e. the overall constant $C_{\rm bc}$  in \eq{omgwbubbles}, cancels in the definition of $R_{ab}(f)$ given in \eq{defRab}. We then compute 
$R_{ab}(f)$ varing just the peak frequency $f_{\rm bc}$ in \eq{omgwbubbles}, taking three illustrative values
$f_{\rm bc}=10, 100$ and 1000~Hz.
The results, at the 3G detector network that we are studying, are shown in \cref{fig:r_ab_nopowerlaw}. We see, once again, that the confusion noise from BBHs is negligible (having set again $\snrth=12$ as threshold for the resolved sources). For the BNSs, the result depends on the spectrum considered, and on the detector pair. For the (${\rm ET}_1$, ${\rm ET}_2$) pair, for instance, 
the BNS confusion background is negligible when $f_{\rm bc}=1000$~Hz.  The effect is instead maximum, among these three cases, for $f_{\rm bc}=10$~Hz. In any case we see that, for all detector pairs and all spectra considered, the values of $S_{\cal N}/N$, even when they are larger than one, still are only of the order of a few, and therefore do not have a dramatic effect on the sensitivity to cosmological searches, at this 3G network.

Once again, we emphasize that these results depend strongly on the detector network considered, and a comparison of the results for different networks is deferred to further work.

We finally observe that we have focused here on the correlated noise induced by the astrophysical background, neglecting correlated instrumental noise. These will also be present, in particular for ET in its triangular configurations, which features  three colocated detectors; in this case,  correlated Newtonian noise could dominate stochastic searches below ${\cal O}(50-100)$~Hz~\cite{Janssens:2022xmo,Branchesi:2023mws,Janssens:2024jln}, or in any case at least up to 30~Hz assuming an optimistic factor of 10 of Newtonian noise cancellation~\cite{Badaracco:2019vjq}. In such frequency range, and for colocated detectors, the correlated noise from astrophysical backgrounds, even if above the nominal PLS computed assuming uncorrelated noise, might in any case be a subleading effect, and the dominant limitation to stochastic searches could come from such correlated instrumental noise.

\section{Conclusions}\label{sect:conclusions}

In this paper we have studied the effect of the astrophysical confusion noise generated by BBHs and BNSs on the search for cosmological stochastic background, at 3G detectors, laying down the conceptual foundations for a first-principle approach, and evaluating the numerical  result for a 3G detector  network and different cosmological signals.

Our strategy has been to follow  a procedure which mimics as much as possible the actual observational process: we have generated an ensemble  of  BBH and BNS signals from a realistic population model and 
for each event, and a given noise realization in the detectors of the network, we have determined whether the event is resolved or unresolved; for the resolved events, we have reconstructed its signal, including again the effect of the noise. Knowing the true signal that we have injected and the result of the reconstruction, we can compute the error in the subtraction process and then, for each detector in the network, we have a ``partially cleaned'' output stream that includes: a given realization of the  instrumental noise; the signal (projected onto each detector)  from the sources that we have injected  and that, in the given noise realization, are unresolved;    the difference between the actual injected signals   of  the resolved sources and the reconstructed ones, again projected onto each detector;
and, possibly, a cosmological signal. We have then 
performed two-detector correlations of these partially cleaned outputs, to extract a possible cosmological signal, 
treating the sum of the instrumental noise and the astrophysical ``confusion noise'' (i.e., the sum of the contribution to the output of unresolved sources plus the error on resolved sources) as an effective noise. The peculiarity of this effective noise is that it is correlated among the detectors in the network (even when the instrumental noise is uncorrelated), because it depends on the common astrophysical signal, as well as on its reconstruction, which involves the whole detector network. We have studied how to characterize  this correlated noise, and we have computed its effect in the search for a cosmological background, averaging over a large number of realizations of the instrumental noise. 

An important point that  we have emphasized, and that was overlooked in previous studies, is that the effect of the astrophysical confusion  noise also depends on the specific cosmological search that is being performed, since this fixes the filter function that is used in a two-detector correlation. An upper bound on the  effect of the astrophysical confusion noise can then be obtained by choosing the filter function that maximizes it. 
We have then seen how to characterize the astrophysical confusion noise when one uses the filter optimized to a given  cosmological  search; in particular,  the information on the relative importance of the astrophysical confusion noise, as a function of frequency, can be described by the  function $R_{ab}(f)$ introduced in \cref{sect:defRab}.

After having laid down this formalism, we have  performed extensive numerical simulations to compute the effect of the astrophysical confusion noise on cosmological searches at a specific 3G network, featuring an ET detector in Europe (taken in its triangular version) and two CE detectors in the US (of 40 and 20~km, respectively).
The corresponding results for the upper bound on the astrophysical confusion noise are shown in 
\cref{fig:subtraction_test_different_snrth} and show that, for BBHs, even the upper bound on the  confusion noise that they generate can be pushed below the floor determined by the instrumental noise while, for  BNS, this upper bound is below the instrumental noise   up to $f\sim 100$~Hz.

The actual effect of the astrophysical confusion noise on a specific cosmological search, however, can only be assessed by performing the two-detector correlation with the filter that is actually used in the given cosmological search, and the upper bounds displayed  in \cref{fig:subtraction_test_different_snrth} can in fact be quite generous, because the optimal filters for the two cases are very different. Indeed, we found in \cref{fig:R_ab} that the actual contribution of the astrophysical confusion noise to given stochastic searches is significantly smaller than the upper bound shown in \cref{fig:subtraction_test_different_snrth}. 
The main conclusion from this numerical part is that, for the specific 3G network considered, the BBHs can be subtracted so that they do not degrade the sensitivity to cosmological backgrounds, but in fact also the BNS background has an effect which is much less important than what suggested by earlier estimates: it does not affect searches at $f\,\gsim\, 100$~Hz, and at lower frequencies only degrades the sensitivity by a factor of order a few. We have emphasized that these numerical  results depends strongly on the 3G network considered. We defer to further work a systematic study of the results for different 3G networks.

\normalsize

\let\oldaddcontentsline\addcontentsline
\renewcommand{\addcontentsline}[3]{}
\begin{acknowledgments}
We thank  Stefano Foffa for many useful discussions.
The research of  F.I., 
M.~Mag. and N.M. is supported by  the  Swiss National Science Foundation, grant 200020$\_$191957, and  by the SwissMap National Center for Competence in Research. E.B. and M.~Mag. are supported by the SNSF
grant CRSII5$\_$213497. 
The work of M.~Man. received support from the French government under the France 2030 investment plan, as part of the Initiative d'Excellence d'Aix-Marseille Universit\'e -- A*MIDEX AMX-22-CEI-02. 
Computations made use of the Baobab cluster at the University of Geneva.
\end{acknowledgments}
\let\addcontentsline\oldaddcontentsline

\appendix

\section{Comparison with the literature}\label{sect:comparison}

In the literature there have been various recent attempts at estimating the residual left by the unresolved sources and by the error on the resolved sources~\cite{Sachdev:2020bkk,Zhou:2022otw,Zhou:2022nmt,Pan:2023naq}. These works postulate that the spectral density of the astrophysical confusion noise has the form
\be\label{ScongOmerrOmunres}
S_{\rm conf}(f)=S_{\rm unres}(f)+S_{\rm err}(f)\, ,
\ee
where $S_{\rm unres}(f)$ is the contribution from unresolved sources and $S_{\rm err}(f)$ from the error in the subtraction of the resolved sources. 
Equivalently,
\be\label{omegacongOmerrOmunres}
\Omega_{\rm conf}(f)=\Omega_{\rm unres}(f)+\Omega_{\rm err}(f)\, ,
\ee
where 
\bees
\Omega_{\rm conf}(f)&\equiv&\frac{4\pi^2}{3H_0^2}\, f^3 S_{\rm conf}(f)\, ,\\
\Omega_{\rm unres}(f)&\equiv&\frac{4\pi^2}{3H_0^2}\, f^3 S_{\rm unres}(f)\, ,\\
\Omega_{\rm err}(f)&\equiv&\frac{4\pi^2}{3H_0^2}\, f^3 S_{\rm err}(f)\, .
\ees
From \eq{eq:Shsum},
\be
S_{\rm unres}(f)= \frac{1}{T}
\sum_{i=1}^{N_{\rm unres}}
\[  |\tilde{h}_{+,i}(f)|^2 + |\tilde{h}_{\times,i}(f)|^2 \] \, ,
\ee
where the sum is over the unresolved sources. For $S_{\rm err}(f)$, different forms have been postulated. 
Ref.~\cite{Sachdev:2020bkk} (SRS) proposes
\be\label{S_SRS}
S^{\rm SRS}_{\rm err}(f) = \frac{1}{T}
\sum_{i=1}^{N_{\rm res}}\sum_{A=+,\times}
|\tilde{h}^{\rm true}_{A,i}(f)-\tilde{h}^{\rm obs}_{A,i}(f)|^2 
\, ,
\ee
where the sum is now over the resolved sources. This expression has also been used in Ref.~\cite{Zhou:2022otw,Zhou:2022nmt}. Different proposals 
for $S_{\rm err}(f)$ have been put forward in Ref.~\cite{Pan:2023naq}. The authors of  Ref.~\cite{Pan:2023naq}  observe that \eq{S_SRS} does not have a clear operational meaning: the output of a detector is given by the combination $h_a(t)$ given by \eq{hFhF}, rather than by the two separate polarizations. So, they rather work in terms of $h_a(t)$ and propose a quantity that depend on the $a$-th detector,
\be\label{S_PYlinear1}
 S^{\rm PY1}_{a, \rm err}(f)=\dfrac{1}{T}
\dfrac{2}{\langle F_+^2\rangle_a + \langle F_\times^2\rangle_a} \sum_{i=1}^{N_{\rm res}} 
|\tilde{h}^{\rm true}_{a,i}(f) -\tilde{h}^{\rm obs}_{a,i}(f) |^2
\, ,
\ee
where the label PY1 indicates that this is the first subtraction procedure suggested in 
\citeauthor{Pan:2023naq}~\cite{Pan:2023naq}.
A second option, again suggested in Ref.~\cite{Pan:2023naq}, and that we will label as PY2, is to consider the quantity
\bees\label{S_PYlinear2}
&&\hspace{-14mm} S^{\rm PY2}_{a, \rm err}(f)=\dfrac{1}{T}
\dfrac{2}{\langle F_+^2\rangle_a + \langle F_\times^2\rangle_a}\nn\\
&&\hspace{3mm}\times \sum_{i=1}^{N_{\rm res}}  \[ 
|\tilde{h}^{\rm true}_{a,i}(f)|^2 - |\tilde{h}^{\rm obs}_{a,i}(f) |^2
\]
\, ,
\ees
where one takes the difference of the modulus squared, rather than the modulus squared of the difference. Further ``refined subtractions'' have been proposed in Ref.~\cite{Pan:2023naq}.

It should be stressed, however,  that \eqss{S_SRS}{S_PYlinear1}{S_PYlinear2}, as well as the further refinements proposed in Ref.~\cite{Pan:2023naq}, are all heuristic proposals for the expression of the residual background after the subtraction, which have not been backed  by any derivation and, in this sense, should be considered at most  
as  guesses for the   residual error left by the subtraction procedure. We also observe that these heuristic approaches give numerical results that are quite different among them, even in order of magnitude, so at most one of them could be a good proxy for the correct procedure.\footnote{In Ref.~\cite{Pan:2023naq} it seems implicit that the ``correct'' subtraction procedure is the one that gives the smallest value for the astrophysical confusion noise. However, even this is not supported by any argument, since none of the above mathematical formulas has any  clear connection with what is actually done when correlating the cleaned outputs of two detectors.}

In our approach, in contrast, we 
have developed  a   procedure that corresponds (at least, in a first approximation) to what will be actually done, observationally,  in a search for cosmological stochastic backgrounds:  we 
first remove the best estimate for the resolved sources from the data stream of each detector, and we then  correlate (with some filter function) the cleaned outputs of different detectors; to add a further degree of realism, we included the dependence of the reconstruction procedure on the specific realization on noise, and we averaged over many noise realizations. 

We will now compare our results with these more heuristic  approaches.
First of all, we observe that there are important structural differences between the results that we obtain and the formulas proposed above.  None of the above formulas has any connection with the actual observational procedure of ``cleaning'' the output of each detector in a network and then correlating the cleaned outputs, and this shows up in the structure of the proposed formulas.
In particular:

\begin{itemize}

\item As we repeatedly stressed, see e.g. the discussion below \eq{SoverNdelnoiseeff}, the signal-to-noise ratio of the astrophysical confusion noise is  not an intrinsic property of the astrophysical populations and of the detector network, but also depends on the cosmological background that we are searching, because the filter function $\tilde{Q}(f)$ in 
\eq{SoverNdelnoiseeff} is the one that optimizes the search for that cosmological signal. Such a dependence is absent in any of the above proposals, so they can at most be compared with the upper bound on the astrophysical  confusion noise obtained choosing the filter function that maximizes its own signal-to-noise ratio in a two-detector correlation,  that we discussed in \cref{sect:Upper}. 

\item Even in that case our result for the spectral density, given by \eqs{defSconf}{eq: correlator_subtraction_full3}, depends on the output of a pair of detectors $(a,b)$ through
($\tilde{h}^{\rm true}_{a,i}$,  $\tilde{h}^{\rm obs}_{a,i}$) and  
($\tilde{h}^{\rm true}_{b,i}$,  $\tilde{h}^{\rm obs}_{b,i}$). In contrast, 
\eq{S_SRS}, as already observed in Ref.~\cite{Pan:2023naq}, does not depends on the projection of the signal on the two detectors, but only on $\tilde{h}^{\rm true}_{A,i}(f)$ and $\tilde{h}^{\rm obs}_{A,i}(f)$, so it cannot match with the results of our computation. On the other hand, \eqs{S_PYlinear1}{S_PYlinear2} are quadratic in the signal  $\tilde{h}^{\rm true}_{a,i}$,  $\tilde{h}^{\rm obs}_{a,i}$, of a single detector, rather than being bilinear quantities in the output of the two detectors, so they cannot reproduce, nor even as a proxy, the result of performing a correlation among two detectors.

\end{itemize}

In any case, it is interesting to see if our results  can be put in a form similar to \eq{ScongOmerrOmunres} and, in case,  what would be the expression for $S_{\rm err}(f)$ derived from our first-principle approach.
To this purpose, let us  rewrite the correlator (\ref{eq: correlator_subtraction_full3}) in a form that distinguishes more explicitly between the contribution of the unresolved sources and that from the error on the resolved sources. This  can be obtained by carrying 
$\sum_{i=1}^{\nev}$ inside the noise averages  in \eq{eq: correlator_subtraction_full3} (since the sum over all events and the noise average commute), and using $\sum_{i=1}^{\nev}\theta_i=\sum_{i=1}^{N_{\rm res}}$, where now the index $i$ only runs over the $N_{\rm res}$ resolved events. Note, however, that the sums over  resolved events, now, cannot be carried back outside  the noise averages, because which events are resolved and which are not, as well as the value of  $N_{\rm res}$, depends in general on the noise realization.
Introducing the notation
\bees
\tilde{h}^{\rm true}_{a} &=&\sum_{i=1}^{\nev}\tilde{h}^{\rm true}_{a,i}\, ,\\
\tilde{h}^{\rm obs}_{a} &=&\sum_{i=1}^{{N_{\rm res}}}\tilde{h}^{\rm obs}_{a,i}\, ,
\ees
we then obtain
\bees
{\neff}_n &=&\big( \tilde{h}^{\rm true}_{a} - \langle  \tilde{h}^{\rm obs}_{a}\rangle_n \big)^*
\big( \tilde{h}^{\rm true}_{b} - \langle  \tilde{h}^{\rm obs}_{b}\rangle_n \big)\nn\\
&&
+\langle  \tilde{h}^{\rm obs,*}_{a}\tilde{h}^{\rm obs}_{b}\rangle_n
-\langle  \tilde{h}^{\rm obs,*}_{a}\rangle_n\,  \langle\tilde{h}^{\rm obs}_{b}\rangle_n
\nn\\
&&-\langle  \tilde{n}^*_a\tilde{h}^{\rm obs}_{b}+ \tilde{n}_b\tilde{h}^{{\rm obs},*}_{a} \rangle_n\, .
\label{neffapp}
\ees
Observe that the separation between resolved and unresolved events depends on the noise realization: particularly for events with a network SNR close to the detection threshold, combining the signals with different realizations of the noise can make them detectable or not. Therefore, by itself $\tilde{h}^{\rm true}_{a}$ does not admit a separation into resolved and unresolved contributions, unless we specify the noise realization. However we can write, trivially,
$\tilde{h}^{\rm true}_{a}= \langle \tilde{h}^{\rm true}_{a} \rangle_n$, since $\tilde{h}^{\rm true}_{a}$ does not depend on the noise realization and, once inside the noise average, it becomes possible to split it into the contribution from resolved and unresolved sources, so 
\be\label{htrueseparation}
\tilde{h}^{\rm true}_{a} =\langle \tilde{h}^{\rm true, res}_{a} \rangle_n
+\langle \tilde{h}^{\rm true, unres}_{a} \rangle_n\, ,
\ee
where
\bees
\tilde{h}^{\rm true, res}_{a}&=&\sum_{i=1}^{N_{\rm res}} \tilde{h}^{\rm true}_{a,i}\, ,\label{hres}\\
\tilde{h}^{\rm true, unres}_{a}&=&\sum_{i=1}^{N_{\rm unres}} \tilde{h}^{\rm true}_{a,i}\, .
\label{unhres}
\ees
In \eq{hres}
the index $i$  runs only over the sources that, in the given noise realization, are resolved, and in \eq{unhres} over the sources that, in the same noise realization, are unresolved. Then, the correlator (\ref{neffapp}) can be split as
\be
{\neff}_n=
{\neff}_n^{\rm unres}+
{\neff}_n^{\rm err}\,,
\ee
where
\be\label{defssunres}
{\neff}_n^{\rm unres}=
\langle\tilde{h}^{\rm true, unres}_{a}\rangle^*_n
\langle\tilde{h}^{\rm true, unres}_{b}\rangle_n
\, ,
\ee
is the contribution from the unresolved sources, while
\bees
{\neff}_n^{\rm err}&=&
\langle\tilde{h}^{\rm true, res}_{a}-\tilde{h}^{\rm obs}_{a}\rangle^*_n
\langle\tilde{h}^{\rm true, res}_{b}-\tilde{h}^{\rm obs}_{b}\rangle_n\nn\\
&&+ \langle\tilde{h}^{\rm true, unres}_{a}\rangle^*_n
\langle\tilde{h}^{\rm true, res}_{b}-\tilde{h}^{\rm obs}_{b}\rangle_n\nn\\
&&+ \langle\tilde{h}^{\rm true, res}_{a}-\tilde{h}^{\rm obs}_{a}\rangle^*_n
\langle\tilde{h}^{\rm true, unres}_{b}\rangle_n \nn\\
&&+\langle  \tilde{h}^{\rm obs,*}_{a}\tilde{h}^{\rm obs}_{b}\rangle_n
-\langle  \tilde{h}^{\rm obs,*}_{a}\rangle_n\,  \langle\tilde{h}^{\rm obs}_{b}\rangle_n
\nn\\
&&-\langle  \tilde{n}^*_a\tilde{h}^{\rm obs}_{b}+ \tilde{n}_b\tilde{h}^{{\rm obs},*}_{a} \rangle_n\,,
\label{neffnerr}
\ees
is the contribution from the error in the reconstruction of the resolved sources. Note that this  includes 
terms which depend on the difference between the actual and the reconstructed signals, as well as terms which are  due to the correlation between the reconstruction of the signal and the noise realization. 

From \eq{defSconf}, we then have
\be\label{defSconfunreserr}
\Seffab(f)\equiv\frac{2}{T}\, \frac{ |
{\neff}_n^{\rm unres}+
{\neff}_n^{\rm err}
|}{|\Gamma_{ab}(f)|}\, .
\ee
It is then convenient to define 
\bees
S^{\rm unres}_{ab}(f)&\equiv&\frac{2}{T}\, \frac{\big|\, \neff^{\rm unres} \big|}{|\Gamma_{ab}(f)|}\, 
\, ,\label{Sunresdef}
\\
S^{\rm err}_{ab}(f)&\equiv&\frac{2}{T}\, \frac{\big|\, \neff^{\rm err} \big|}{|\Gamma_{ab}(f)|}\, 
\, ,\label{Serrneff}\\
S^{\rm cross}_{ab}(f)&\equiv&\frac{2}{T}\, \frac{1}{|\Gamma_{ab}(f)|}
\, \label{defScross}\\
&&\hspace*{-9mm}\times \Big[\,  \big|\, \neff^{\rm unres}+
\neff^{\rm err} \big|\nn\\
&&\hspace*{-5mm}
-\big|\, \neff^{\rm unres} |
-| \neff^{\rm err} \big|\, 
\Big] \, ,\nn
\ees
so that 
\be\label{Sconfthreeterms}
\Seffab(f)=S^{\rm unres}_{ab}(f)+S^{\rm err}_{ab}(f)+S^{\rm cross}_{ab}(f)\, .
\ee
Observe that $S^{\rm cross}_{ab}(f)$  is smaller or equal than zero, because of the Cauchy-Schwarz inequality.

From \eq{defOconf} we  can define the corresponding quantities 
\bees
\Omega^{\rm unres}_{ab}(f) &\equiv&
\frac{4\pi^2}{3H_0^2}\, f^3S^{\rm unres}_{ab}(f)\, ,\label{defOmegaunres}\\
\Omega^{\rm err}_{ab}(f) &\equiv&
\frac{4\pi^2}{3H_0^2}\, f^3S^{\rm err}_{ab}(f)\, ,\\
\Omega^{\rm cross}_{ab}(f) &\equiv&
\frac{4\pi^2}{3H_0^2}\, f^3S^{\rm cross}_{ab}(f)\, .
\ees
Note that, while, as we will show below,  $\Omega^{\rm unres}_{ab}(f)$ is indeed  the energy density (normalized to $\rho_c$) of the unresolved sources, $\Omega^{\rm err}_{ab}(f)$ and $\Omega^{\rm cross}_{ab}(f)$ do not have an interpretation in terms of energy densities. As for  $\Omega^{\rm err}_{ab}(f)$, there is no energy associated to the fact of making an error in the reconstruction of a signal, while $\Omega^{\rm cross}_{ab}(f)$ is even negative. These are simply quantities that characterize the correlated outputs of a pair of detectors, transformed into units that make easy the comparison with the normalized energy density of a cosmological background.

To make contact with \eq{omegacongOmerrOmunres}, let us see how the above expressions simplify if we neglect the fact that the actual SNR of a CBC signal in a detector is determined by how the GW signal combines with a specific noise realization, and we simply use, for the SNR of the $i$-th signal in the $a$-th detector, the standard expression for the optimal SNR (see, e.g., Eq.~(7.51) of Ref.~\cite{Maggiore:2007ulw})
\be\label{SNRVol1}
{\rm SNR}_{a,i}^2 =4\int_0^{\infty}d f\, \frac{ |\tilde{h}_{a,i}(f)|^2}{S_{n,a}(f)}\, ,
\ee
which takes into account the noise only through its average properties encoded in its power spectral density, rather than  depending on the specific realization of the noise at the time of arrival of the signal. In this case, whether a signal goes or not above a threshold $\snrth$ is just a property of the signal (for given detector noise power spectral density). Therefore, the separation between resolved and unresolved sources 
loses any dependence on noise realization and \eq{htrueseparation} simply becomes
\be\label{htrueseparation2}
\tilde{h}^{\rm true}_{a} =\tilde{h}^{\rm true, res}_{a} 
+ \tilde{h}^{\rm true, unres}_{a} \, ,
\ee
while \eq{defssunres} becomes
\bees\label{defssunres2}
\nab_n^{\rm unres}&=&
\tilde{h}^{\rm true, unres}_{a}(f)
\tilde{h}^{\rm true, unres}_{b}(f)\nn\\
&&\hspace*{-20mm}=\sum_{i,j=1}^{N_{\rm unres}} \tilde{h}^{\rm true, *}_{a,i}(f)\tilde{h}^{\rm true}_{b,j}(f)
\, ,
\ees
where we have reinstated explicitly the dependence on the frequency.
Let us now perform the average over ``Universe'' realizations  on this expression.
This is the same as the average already performed in 
\eq{sscor}, where however 
in \eq{tildeaFAastro}
$\nev$ is replaced by $N_{\rm unres}$. Therefore 
\be\label{sscorastro2}
\langle\tilde{h}^{\rm true, unres, *}_a(f)\tilde{h}^{\rm true, unres}_b(f)\rangle_{U}= \frac{T}{2} S_h^{\rm unres}(f) \Gamma_{ab}(f)\, ,
\ee
where 
\be\label{Shunres}
S_h^{\rm unres}(f) = \frac{1}{T}
\sum_{i=1}^{N_{\rm unres}}
\[  |\tilde{h}_{+,i}(f)|^2 + |\tilde{h}_{\times,i}(f)|^2 \] \, .
\ee
Therefore, in the approximation in which the SNR is only determined by the noise spectral density, and not by the specific noise realization, 
\be\label{nabunres}
\nab^{\rm unres}=
\frac{T}{2} S_h^{\rm unres}(f) \Gamma_{ab}(f)\, .
\ee
Plugging this into \eq{Sunresdef}, we then find that $\Gamma_{ab}(f)$ cancels and
\bees
S^{\rm unres}_{ab}(f) &=& S_h^{\rm unres}(f) \label{Snabunres}\\
&=&  \frac{1}{T}
\sum_{i=1}^{N_{\rm unres}}
\[  |\tilde{h}_{+,i}(f)|^2 + |\tilde{h}_{\times,i}(f)|^2 \] \, ,\nn
\ees
and, correspondingly, $\Omega^{\rm unres}_{ab}(f)$ becomes the same as
$\Omega_{\rm gw}^{\rm unres}(f)$, where
\bees
\Omega_{\rm gw}^{\rm unres}(f) &\equiv& \frac{4\pi^2}{3H_0^2}\, f^3 S_h^{\rm unres}(f)\label{Onabunres}\\
&=&\frac{4\pi^2}{3H_0^2}\, f^3\, 
\frac{1}{T}
\sum_{i=1}^{N_{\rm unres}}
\[  |\tilde{h}_{+,i}(f)|^2 + |\tilde{h}_{\times,i}(f)|^2 \] \, .\nn
\ees
We see that, in this approximation, $S^{\rm unres}_{ab}(f)$ loses any dependence from the detector pair $(a,b)$, and is just the spectral density of the unresolved sources. Correspondingly, $\Omega^{\rm unres}_{ab}(f)$ becomes the same as $\Omega_{\rm unres}(f)$, the energy fraction carried by the unresolved sources, again independently of the detector pair considered.  The noise averages in  \eq{defssunres}, that have been neglected in the derivation of  \eq{nabunres} [and therefore
of \eqs{Snabunres}{Onabunres}] further
take into account that, depending on the specific noise realization, sources for which the SNR, computed with \eq{SNRVol1}, is close to the threshold value, could be pushed above or below the threshold by the specific noise fluctuations at the time of arrival of the event.

We therefore recovered  the term $S_{\rm unres}(f)$ in \eq{ScongOmerrOmunres}, at least in the approximation of neglecting the dependence on the noise realization. From \eq{Sconfthreeterms} we see that, in our first-principle computation, there are two more terms, $S^{\rm err}_{ab}(f)$ and $S^{\rm cross}_{ab}(f)$;  the latter  is a mixed term involving both $\neff^{\rm unres}$ and $\neff^{\rm err}$. However, when one of the two correlators, $\neff^{\rm unres}$ or $\neff^{\rm err}$, is much larger than the other, this term goes to zero. We can then identify
$S^{\rm err}_{ab}(f)$  with the first-principle expression for the contribution to the spectral density from the error on the resolved sources, in a two-detector correlation, and when one chooses the filter function that maximizes the astrophysical confusion noise. From \eqs{neffnerr}{Serrneff}, we see that it is sensibly more complex than any of the expression proposed in \eqss{S_SRS}{S_PYlinear1}{S_PYlinear2}, and has a rather different structure,  as a sum of terms made with products of a quantity relative to detector $a$ and one to detector $b$, each one averaged over noise realizations. So, none of the heuristic proposals in the literature for $S^{\rm err}_{ab}(f)$
can really be justified, even a posteriori, by our first-principle approach.

It is therefore not surprising that our results are, numerically, quite different from those of Refs.~\cite{Zhou:2022otw,Zhou:2022nmt} and of Ref.~\cite{Pan:2023naq} (which are also quite different among them; as shown in Ref.~\cite{Zhou:2022otw,Zhou:2022nmt}, one cannot compare with 
the results in Ref.~\cite{Sachdev:2020bkk} because
the parameter space of the waveform used there  was too small, resulting in an order-of-magnitude overestimate of the quality of the subtraction). 
We also find other significant differences in the numerical results. In particular, 
as discussed in \cref{sect:rescorr}, the dependence of our results on $\snrth$ is monotonic. The lower we choose $\snrth$, i.e. the more sources we resolve, the lowest is the overall confusion noise from unresolved sources plus the error on resolved sources. At most, this effect can saturate when we include resolved sources that are too poorly constrained, as we see indeed in \cref{fig:subtraction_test_different_snrth}.
In contrast, Refs.~\cite{Zhou:2022otw,Zhou:2022nmt}  find that, as $\snrth$ is lowered from large values,   the sum of the confusion noise from unresolved sources plus the error on resolved sources at first decreases, as we also find, but then eventually, below some critical value  $(\snrth)_c$, starts increasing again, so that resolving more sources  results in an increase of the astrophysical confusion noise. For BNSs, Refs.~\cite{Zhou:2022otw,Zhou:2022nmt} find that $(\snrth)_c\simeq 20$, while no such effect is seen in our results.  For BBHs, they find that the effect (in most of the frequency range) sets in around $(\snrth)_c\simeq 10$; furthermore, this critical value of the SNR found in Refs.~\cite{Zhou:2022otw,Zhou:2022nmt} is frequency dependent, and  for BBHs, at $f=1$~kHz,  they get  $(\snrth)_c\simeq 30$; again, none of these effects are seen in our BBH results where, for all frequencies, the behavior with $\snrth$ is monotonic (we have checked this down to  $\snrth=8$), and around 1~kHz
the various curves saturate to a common curve (within the statistical fluctuations inherent to our average over many noise realizations), but do not cross each other.
The effect found in Refs.~\cite{Zhou:2022otw,Zhou:2022nmt} would imply that, when a CBC is detected at low SNR, and therefore in general is reconstructed with a large error on its parameters, it is better not to use at all this information, and treat the source as unresolved, rather than subtracting it from the detectors' outputs. However, 
unless one is introducing a bias (and our maximum likelihood estimator is unbiased)
having some information cannot be worse than having no information at all. At worst, the information can be basically useless, if the error on the parameters of the waveform is too large, and therefore does not add anything; we should then expect that, lowering $\snrth$ below some value,  the effectiveness of the subtraction procedure will saturate, which is indeed what we find. We believe that the effect found in Refs.~\cite{Zhou:2022otw,Zhou:2022nmt}, and the corresponding ``optimal value of $\snrth$'',  is  an artifact of  their use of the Fisher matrix formalism (in contrast to our exact maximization of the likelihood) that, especially at low $\snr$, can occasionally produce very large errors (and/or, possibly,  of their subtraction procedure which, as discussed above, is based on a heuristic expression for $\omerr$ that is not rooted in any first-principle computation). Indeed, while in generic population studies such as those in Ref.~\cite{Branchesi:2023mws}, getting a wrong parameter estimation on a few events over a large population of $10^5$ detections has little or no consequences, in the present case, where we are interested precisely in the estimate of the error in the reconstruction, a few outliers due to ill-conditioned Fisher matrices can have severe consequences (see e.g. \cref{footnote:FIM_fail_dLi}). 

\let\oldaddcontentsline\addcontentsline
\renewcommand{\addcontentsline}[3]{}
\bibliography{bibliography}
\let\addcontentsline\oldaddcontentsline

\end{document}